\documentclass[prx,aps,amssymb,twocolumn,reprint,superscriptaddress,longbibliography]{revtex4-1}
\usepackage{amsmath}
\usepackage{amssymb}
\usepackage{amsthm}
\usepackage{amsfonts}
\usepackage{listings}
\usepackage{enumerate}
\usepackage{latexsym}
\usepackage{color}
\usepackage{xcolor}
\usepackage{bm}
\usepackage{hyperref}
\hypersetup{
 pdfnewwindow=true, colorlinks=true,
 linkcolor=blue, anchorcolor=blue,
 citecolor=blue, filecolor=blue,
 menucolor=blue, urlcolor=blue}

\newcommand{\MST}{${\rm Mn_3Si_2Te_6}$}
\usepackage{psfrag}

\usepackage{graphicx}
\usepackage{subfigure}
\usepackage{lipsum}
\usepackage{float}
\usepackage{array} 
\usepackage{soul} 
\usepackage{titlesec}

\newcommand*{\dif}{\mathop{}\mathrm{d}}

\begin{document}
\tolerance 10000

\draft

\title{A Framework for Understanding Colossal Magnetoresistance and Complex Resistivity Behaviors in Magnetic Semiconductor}

\author{Zhihao Liu}
\affiliation{Beijing National Laboratory for Condensed Matter Physics and Institute of Physics, Chinese Academy of Sciences, Beijing 100190, China}
\affiliation{University of Chinese Academy of Sciences, Beijing 100049, China}
 \author{Zhong Fang} 
\affiliation{Beijing National Laboratory for Condensed Matter Physics and Institute of Physics, Chinese Academy of Sciences, Beijing 100190, China}
\affiliation{University of Chinese Academy of Sciences, Beijing 100049, China}
\affiliation{Songshan Lake Materials Laboratory, Dongguan, Guangdong 523808, China}
\author{Hongming Weng}
\affiliation{Beijing National Laboratory for Condensed Matter Physics and Institute of Physics, Chinese Academy of Sciences, Beijing 100190, China}
\affiliation{University of Chinese Academy of Sciences, Beijing 100049, China}
\affiliation{Songshan Lake Materials Laboratory, Dongguan, Guangdong 523808, China}
\author{Quansheng Wu}
\email{quansheng.wu@iphy.ac.cn}
\affiliation{Beijing National Laboratory for Condensed Matter Physics and Institute of Physics, Chinese Academy of Sciences, Beijing 100190, China}
\affiliation{University of Chinese Academy of Sciences, Beijing 100049, China}

\begin{abstract}
Colossal magnetoresistance (CMR) is commonly observed in magnetic semiconductors when an external magnetic field is applied, usually accompanied by anomalous resistivity peaks or humps which were previously considered as evidence of a metal-insulator transition (MIT). Previous research efforts primarily focused on elucidating the CMR effect in ferromagnetic semiconductor systems, which may be inapplicable to antiferromagnetic or ferrimagnetic systems. In our work, we propose a framework to unravel the mechanisms underlying the CMR related phenomenon in magnetic semiconductor, and apply it to the ferrimagnetic semiconductor \MST. A two-carrier model, with electron and hole carriers generated by thermal excitation across the magnetization-dependent band gap, is employed, which accurately reproduces the observed $\rho(B, T)$ curves. Additionally, the shift of $T_c$ (or the anomalous resistivity peak) with increasing direct currents, previously attributed to current control of the chiral orbital current (COC) state, is also reproduced within our framework by properly accounting for the Joule heating effects. Our work provides a quantitative methodology for analyzing and calculating the novel resistivity curves in magnetic semiconductors and clarifies the controversy surrounding the origins of CMR and the concomitant complex behaviors.
\end{abstract}

\maketitle

The discovery of colossal magnetoresistance (CMR), characterized by dramatic decline in electrical resistivity under an external magnetic field, has emerged as a cornerstone phenomenon in condensed matter physics and materials science, with profound implications for sensitive magnetic field detector or reliable magnetic storage devices. Initially studied extensively in the doped perovskite manganites such as $\rm La_{1-x}Ca_xMnO_3$ systems \cite{jinThousandfold1994,jinColossal1994,tokuraGiant1994,asamitsustructural1995,urushibaraInsulatormetal1995, roderLattice1996, ramirezColossal1997}, CMR and the anomalous resistivity peak observed near Curie temperature $T_c$ were conventionally attributed to the double exchange (DE) mechanisms \cite{zenerInteraction1951, andersonConsiderations1955,degennesEffects1960}. Later the DE model was criticized and modified through the incorporation of the Jahn-Teller (JT) effect which arises from the lattice distortion \cite{millisDouble1995, millisDynamic1996, salamonphysics2001}. However, CMR and the resistivity peak are so prevalent in ferromagnetic semiconductors that they are observed even in systems lacking the DE or JT effect like EuS, EuO, $\rm CdGr_2S_4$, $\rm HgCr_2Se_4$ \cite{shapiraResistivity1972, oliverConductivity1972, shapiraEuO1973, konnoElectrical1998, sunColossal2010, linSpin2016a}. For this reason, Nagaev proposed a magnetoimpurity theory that attributes CMR to the enhancement of local magnetic moments near impurity clusters due to spatial fluctuations in magnetization and electron density \cite{nagaevMagnetoimpurity1999, nagaevColossalmagnetoresistance2001}. Additional works following the polaron-based framework include magnetic polaron model, which describes the formation of nonoverlapping polaron regions at low carrier density and strong electron-core spin coupling \cite{majumdarMagnetoresistance1998, sullowMagnetotransport2000, yangMagnetic2004, pohlitEvidence2018}. Some studies focused on the polaron or charge correlations, proposing that the competition between the ferromagnetic state and the short-range correlated charge-ordered antiferromagnetic state by incorporating the AF superexchange $J_{\rm AF}$, serves as the origin of CMR \cite{tomiokaMagneticfieldinduced1997,yunokiStatic1998,alonsoDiscontinuous2001, dagottoColossal2001, tokuraCritical2006, senCompeting2007}. It is evident that research into the origins of the CMR phenomenon is marked by a diversity of opinions and has yet to be fully understood.

The aforementioned studies have been developed within the framework of ferromagnetic exchange models and are applicable to materials with a ferromagnetic ground state. Nevertheless, recent studies have revealed that CMR and the anomalous resistivity peak were also observed in antiferromagnetic or ferrimagnetic semiconductors such as $\rm HgCr_2S_4$, $\rm Eu_5In_2Sb_6$, $\rm EuP_3$, \MST, $\rm EuTe_2$, $\rm EuSe_2$, $\rm EuMnSb_2$, ${\rm Eu}T_2X_2$ ($T$ = Cd, Zn and $X$= P, As), etc. \cite{weberColossal2006, rosaColossal2020, wangFieldInduced2020, niColossal2021,yinLarge2020,yangColossal2021,takeuchiFieldinduced2024, dongSimultaneous2024, yinMagnetism2024, wangColossal2021, krebberColossal2023, luoColossal2023, usachovMagnetism2024}, which poses a challenge to the previously established framework. A notable discrepancy is the significant deviation of the resistivity peak location from the antiferromagnetic phase transition point, which starkly contrasts with the behavior observed in ferromagnetic semiconductors. In reality, the emergence of resistivity peaks under applied magnetic fields does not universally serve as a definitive indicator of metal-insulator transitions (MIT) \cite{zhanginadequacy2025}. Other complex resistivity behaviors include the resistivity peak movement toward higher temperatures with increasing magnetic fields, or the peak shift toward lower temperatures in \MST\ with increasing currents\cite{zhangControl2022}. A coherent theoretical framework capable of  systematically explaining or reproducing these observations remains elusive.

In this work, we propose a framework for magnetic semiconductors that bridges these challenges by integrating a two-carrier model with magnetization-dependent bandgap engineering and thermal equilibrium analysis. By explicitly accounting for the thermal excitation of electron-hole pairs across a dynamically modulated band gap, our model is capable of quantitatively calculate the complex resistivity behaviors. We apply our method to the highly focused ferrimagnetic semiconductor \MST, yielding results that are in excellent agreement with experimental observations including the CMR, the anomalous resistivity peaks (or humps) as well as their shift.  Crucially, we also demonstrate that Joule heating, rather than COC states, can drives the current-induced suppression of $T_c$ and the observed first-order phase transition signal in resistivity. This framework is further applied to the antiferromagnetic semiconductor $\rm EuSe_2$ and ferromagnetic semiconductor $\rm La_{1-x}Ca_xMnO_3$.

\textit{\textbf{Two-carrier model with dynamically modulated bandgap.}} Our theoretical framework is formulated within a thermally excited two-carrier model based on the first-principles calculated energy gap. The investigation of magnetotransport in semiconductors involve two sequential processes: (1) electron-hole carriers are generated across the bandgap due to thermal activation, and (2) these carriers, under the influence of the Lorentz force, undergo motions governed by Drude theory. Critically, the bandgap may exhibit a strong magnetization dependence,  thereby undergoing dynamical modulation by the magnetic field and  temperature. Dynamic bandgap modulation combined with Lorentz force effect may lead to unconventional transport phenomena.

The carrier concentrations of  thermally excited electron-hole carriers are determined using the following expressions:

\begin{gather}
    n_e=\int_{E_{\rm CBM}}^\infty D_e(\epsilon)f(\epsilon-\mu, T)\dif\epsilon, \label{density_e}\\
    n_h=\int_{-\infty}^{E_{\rm VBM}} D_h(\epsilon)\left(1-f(\epsilon-\mu, T)\right)\dif\epsilon.
\end{gather}

$E_{\rm CBM}$ and $E_{\rm VBM}$ represent the energy values at the conduction band minimum (CBM) and the valence band maximum (VBM), $f(\epsilon-\mu, T)$ is the Fermi distribution function with $\mu$ representing the chemical potential, and $D_e(\epsilon)$ and $D_h(\epsilon)$ are the densities of states of the electron and hole carriers, expressed as follows:
\begin{gather}
D_e(\epsilon)=C_e(\epsilon- E_{\rm CBM})^{1/2}, C_e=4\pi(\frac{2m_e}{h^2})^{3/2},\\
D_h(\epsilon)=C_h(E_{\rm VBM}-\epsilon)^{1/2}, C_h=4\pi(\frac{2m_h}{h^2})^{3/2},
\end{gather}
where the quadratic band dispersion approximation has been employed: $\epsilon_e(k)=\frac{\hbar^2k^2}{2m_e}+ E_{\rm CBM}$, $\epsilon_h(k)=-\frac{\hbar^2k^2}{2m_h}+ E_{\rm VBM}$. Since the electron and hole carriers are excited in pairs in a semiconductor, their concentrations are considered to be the same. Applying equation $n_e=n_h$, the carrier concentrations and the chemical potential can be determined simultaneously.

$E_{\rm CBM}$ and $E_{\rm VBM}$ may be magnetization-dependent, as the out-of-plane magnetization ($M \parallel c)$ can lift the degeneracy of the spin-polarized bands and narrow the energy gap, where the reduction in the bandgap is proportional to the $z$ component of magnetization \cite{seoColossal2021}. Thus the following expressions are used to describe such a change in the band gap:
\begin{gather}
    E_{\rm CBM}=\Delta_{\rm gap}/2-G\left(M(B, T)\right),\label{E_CBM}\\
    E_{\rm VBM}=-\Delta_{\rm gap}/2+G\left(M(B, T)\right),
\end{gather}
where $\Delta_{\rm gap}$ is the ground state gap value, $M(B, T)$ is the field- and temperature-dependent magnetization, and $G$ is a linear magnetization-dependent function, expressed as
\begin{gather}
    G\left(M(B, T)\right)=\frac{1}{2}\Delta E \times M(B, T)/M_s.\label{G_fun}
\end{gather}

$\Delta E$ represents the energy change in band gap between the initial and saturated state, and $M_s$ is the saturated magnetization. In the case of an insulator-to-metal transition occurring, $\Delta E$ is larger than $\Delta_{\rm gap}$.
The carrier concentrations at a given field and temperature are fully determined by Eq. (\ref{density_e}-\ref{G_fun}). The relaxation times $\tau_e$ and $\tau_h$, which determine the carrier mobilities via the relations $\mu_e(T)=e\tau_e(T)/m_e$ and $\mu_h(T)=e\tau_h(T)/m_h$, are assumed to be inversely proportional to the temperature, which works well for most cases \cite{liuCombined2024}:
\begin{gather}
    \tau(T)=\frac{1}{\zeta+\chi T}, 
\end{gather}
where $\zeta$ characterizes the impurity scattering rate, with larger values of $\zeta$ corresponding to higher impurity concentrations, and $\chi$ characterizes the electron-phonon scattering strength. The resistivity curve $\rho(B, T)$ can now be simulated using the two-carrier Drude model as follows:
\begin{gather}
    \rho(B, T)
    =\frac{1}{e}\frac{(n_e \mu_e+n_h \mu_h)+(n_e \mu_e\mu_h^2+n_h\mu_h\mu_e^2)B^2}
    {(n_e\mu_e + n_h\mu_h)^2+
    (n_e - n_h)^2\mu_e^2\mu_h^2B^2}.\label{rxx}
\end{gather}

This framework is universal for calculating the resistivity of magnetic semiconductor materials. The different magnetization curves lead to specific dependencies of the band gap on magnetic field, which may give rise to a diverse set of resistivity curves, such as CMR and anomalous resistivity peaks or humps. Extending the two-carrier model to a multiple-carrier scenario is also straightforward by imposing $\sum_i n_{e,i}=\sum_j n_{h,j}$ and replacing Eq. (\ref{rxx}) with the inverse of the sum of conductivities from all carriers.

\begin{figure*}[htb]
    \centering
    \includegraphics[width=0.95\textwidth]{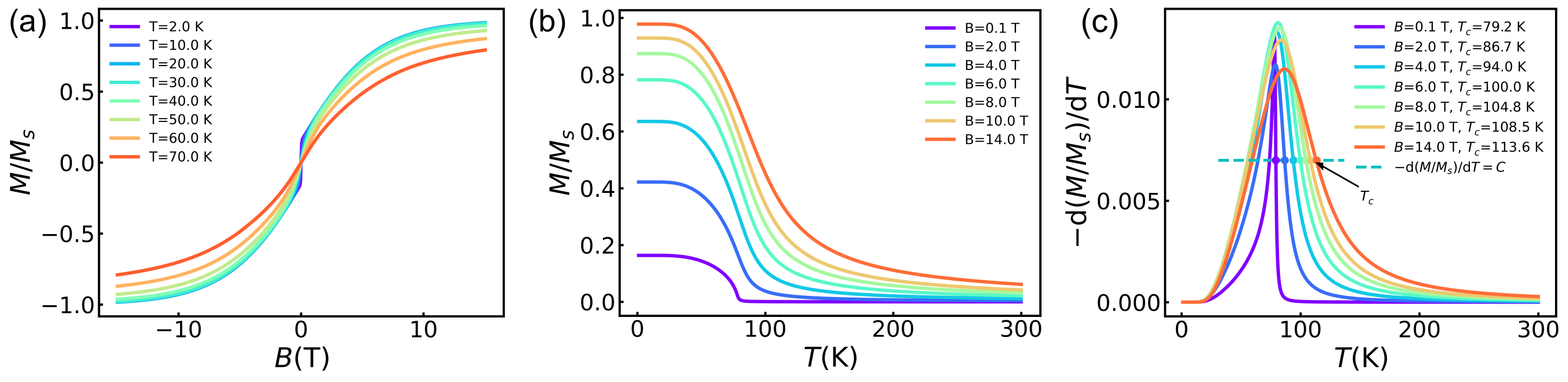}
    \caption{(a) Magnetic field-dependent $M/M_s$ curves at different temperatures. (b) Temperature-dependent $M/M_s$ at different magnetic fields. (c) The rate of decline in magnetization with temperatures, expressed as $-\dif (M/M_s)/\!\dif T$. The intersection range between the dashed cyan line with value of $C$ and the $-\dif (M/M_s)/\!\dif T$ curves determines the intermediate temperature region, with the endpoint of this region labeled as $T_c$.}
    \label{fig:magnetization}
\end{figure*}

\begin{figure*}[htb]
    \centering
    \includegraphics[width=0.95\textwidth]{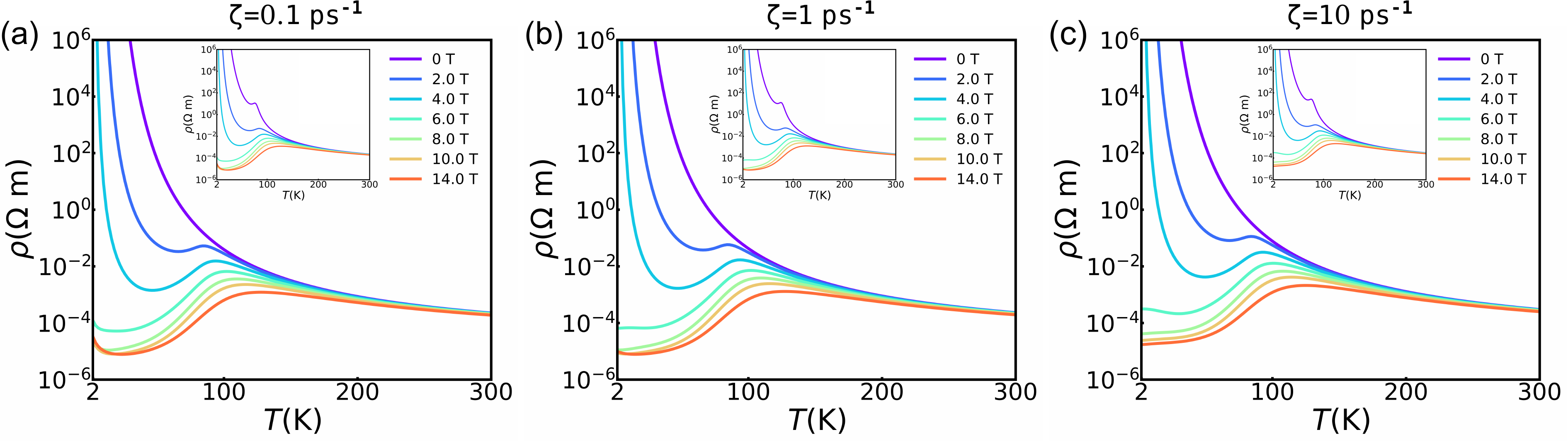}
    \caption{The simulated resistivity curves $\rho(B, T)$, accurately reproduce the observed CMR and resistivity humps in \MST. The impurity concentration related parameter $\zeta$ is sequentially increased from 0.1 to 10 $\rm ps^{-1}$ in (a)-(c), which controls the impurity scattering rate. As $\zeta$ increases, $\rho(B, T)$ exhibits an evolution from a ``semiconducting" decreasing behavior to ``metallic" increasing behavior at low temperature region under high fields. The movement of resistivity hump location toward high temperatures with increasing fields is also observed, almost identical to the location of $T_c$ defined in Fig. \ref{fig:magnetization} (c). Inset: $\rho(B, T)$ curves when considering scattering enhancement near the critical  temperature. }
    \label{fig:rhoBT}
\end{figure*}

\textit{\textbf{Application to \bm \MST.}} The ferrimagnetic semiconductor \MST\  has drawn significant attention in recent years for its drastic negative CMR up to nine orders of magnitude, promising for realizing extremely sensitive spintronic devices. Additionally, as the temperature approaches $T_c$, the increase in resistivity scales with the magnetic field strength, reaching up to two orders of magnitude at high fields, forming large anomalous resistivity humps. Another strange finding is that, for samples from different research groups, the resistivity curves $\rho(T)$ in the low-temperature region under high fields exhibit distinct behaviors. For some samples, $\rho(T)$ continues to decrease with temperature, maintaining a ``semiconducting" characteristics \cite{niColossal2021}, while for others, the curve turns into a slowly increasing, flat curve, displaying a ``metallic" behavior \cite{seoColossal2021, zhangControl2022, zhangCurrentsensitive2024}. An even more interesting observation is the current control of resistivity, where $T_c$ and resistivity are significantly suppressed with the application of increasing direct current. These unusual phenomena have not been thoroughly and precisely described in previous works, whereas our study places them within an unified framework and provides a quantitative explanation.

By constructing appropriate functions (Supplemental Material \cite{Supp}), we can reproduce magnetization curves  that closely match the experimental results with magnetic field applied along $c$ axis, as shown in Fig. \ref{fig:magnetization}(a), (b). Fig. \ref{fig:magnetization}(a) shows the magnetic field-dependent $M/M_s$ at different temperatures. Note that $M$ approaches $M_s$ for $B>14$ T, and at low temperatures, the magnetization undergoes a rapid increase over a small range of magnetic field, consistent with the experimental observations. Fig. \ref{fig:magnetization}(b) shows the temperature-dependent $M/M_s$ at different magnetic fields. We find that as the temperature increases, the rate of magnetization decline accelerates, and after a certain temperature, the rate of decline slows down. Fig. \ref{fig:magnetization}(c) shows the curves representing the rate of decline at different fields, expressed as $-\dif (M/M_s)/\!\dif T$. As the magnetic field increases, the broadening of $-\dif (M/M_s)/\!\dif T$ increases, resulting in the movement of $T_c$ toward high temperatures, which will be discussed later.

Using the magnetization curves constructed above and applying Eq. (\ref{density_e}-\ref{rxx}), we can simulate the resistivity curves $\rho(B, T)$. The energy gap value, $\Delta_{\rm gap}$ of 120 meV, is adopted from first-principles calculations \cite{seoColossal2021}, and $\Delta E$ is set to 160 meV. The effective mass is assumed to be equal to the rest mass of a free electron. $\chi$ is set to $\rm ps^{-1}\rm K^{-1}$ and $\zeta$ is variable. Fig. \ref{fig:rhoBT} shows the obtained resistivity curves. In Figs. \ref{fig:rhoBT}(a)-(c), we sequentially increase the parameter $\zeta$, which controls the impurity scattering rate, from 0.1 to 1, and then to 10 $\rm ps^{-1}$. As $\zeta$ increases, the behavior of $\rho(B, T)$ at low temperatures under high fields evolves from a ``semiconducting" decreasing behavior to a ``metallic" increasing behavior. This may explain the distinct resistivity behaviors observed in samples from different research groups, which are likely attributed to impurity differences. CMR is also observed, with the resistivity dropping by more than ten orders of magnitude up to 14 T. Additionally, the anomalous resistivity humps are reproduced at non-zero magnetic fields, without considering any scattering enhancement near the critical temperature, suggesting that the scattering enhancement may not be the dominant factor. 

% \begin{table}[h]
%         \caption{parameters of our two-carrier model}
% 	\begin{tabular}{|>{\centering}p{1.5cm}|>{\centering}p{1.8cm}|>{\centering}p{1.8cm}|p{1.8cm} <{\centering}|}\hline
% 	&$m$(kg) & $\zeta$($\rm ps^{-1}$) & $\chi$($\rm ps^{-1}\rm K^{-1}$)\\ \hline
%         electron&$9.11\times 10^{-31}$ & $t\times 0.1$ & 0.1\\ \hline
%         hole&$9.11\times 10^{-31}$ & $t\times 0.1$ & 0.1\\
%         \hline
% 	\end{tabular} \label{para}
% \end{table}

In our framework, the primary cause of the anomalous humps is attributed to the specific temperature dependence of the band gap, which is governed by the temperature-dependent magnetization, as described in Eq. (\ref{E_CBM}-\ref{G_fun}). We take the $\rho(B, T)$ curve at 4 T in Fig. \ref{fig:rhoBT}(a) as an example to illustrate this phenomenon, in combination with the magnetization decline rate shown in Fig. \ref{fig:magnetization}(c). In the low-temperature region, $-\dif (M/M_s)/\!\dif T$ is small, indicating a slow decline in magnetization. The widening of the magnetization-dependent band gap is insignificant, and consequently the increase in carrier concentrations due to thermal excitation dominates, resulting in a decrease in resistivity. In the intermediate temperature region, $-\dif (M/M_s)/\!\dif T$ becomes large, suggesting the acceleration of the decline in magnetization. The thermal excitation-induced increasing trend in carrier concentrations is suppressed by the decreasing trend caused by the gap widening, leading to an increase in resistivity. As the temperature increases further, $-\dif (M/M_s)/\!\dif T$ becomes small again and the band gap widening slows down, thus the resistivity resumes its typical semiconductor behavior and exhibits a decrease again. We assume a certain value $C$ such that the width of the intermediate temperature region is determined by $-\dif (M/M_s)/\!\dif T \geq C$, as shown in Fig. \ref{fig:magnetization}(c). As the magnetic field increases, the intermediate temperature region widens, consistent with the expansion of the temperature range where resistivity increases with increasing magnetic fields, as shown in Figs. \ref{fig:rhoBT}(a)-(c). The endpoint of the intermediate temperature region, labeled as $T_c$, which marks the peak value of the anomalous resistivity hump, moves toward higher temperatures with increasing fields, in agreement with the experimental findings. The insets in Figs. \ref{fig:rhoBT}(a)-(c) show the $\rho(B, T)$ curves when the scattering enhancement near the critical temperature is considered, using the phenomenological description (Supplementary Material \cite{Supp}). The resistivity curve at zero field now acquires a small anomalous peak that resembles the experimental observation, while the other curves at non-zero fields show no significant changes. It is noteworthy that this framework is also applicable to a wide range of antiferromagnetic and ferromagnetic materials, such as $\rm EuSe_2$, $\rm EuTe_2$ \cite{dongSimultaneous2024, yangColossal2021}, and $\rm La_{1-x}Ca_xMnO_3$ detailed in the Supplementary Materials \cite{Supp}, demonstrating its universality across distinct magnetic phases.

\begin{figure}[htb]
    \includegraphics[width=0.47\textwidth]{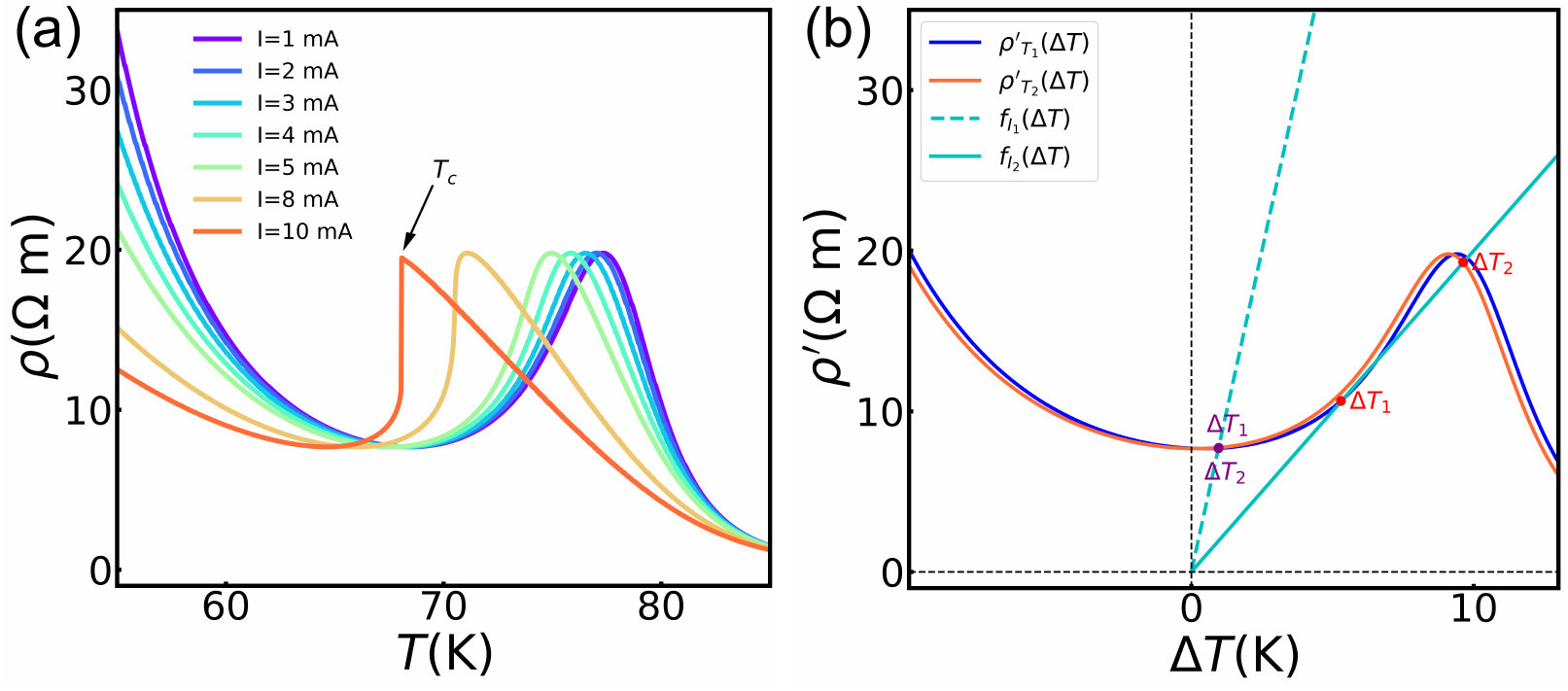}
    \caption{(a) The shift of $T_c$ in resistivity with increasing currents from 1 mA to 10 mA. A first-order-like transition signal in resistivity, characterized by a sharp rise near $T_c$ with large current, is reproduced. (b) Schematic illustration of the shift of $T_c$ and the sharp rise in resistivity. The blue curve $\rho'_{T_1}(\Delta T)$ and the red curve $\rho'_{T_2}(\Delta T)$ are obtained by shifting the resistivity $\rho(T)$ to the left by $T_1$ and $T_2$. The dashed cyan line $f_{I_1}(\Delta T)$ and the solid cyan line $f_{I_2}(\Delta T)$ are the functions defined in the left-hand side of Eq. (\ref{eqn:Tcsolve}) with different current values $I_1$ and $I_2$. $\Delta T_1$ and $\Delta T_2$ are the solutions of temperature deviation from measured environmental temperature originated from Joule heating effect.}
    \label{fig:peakshift}
\end{figure}

\textit{\textbf{Joule Heating Effect:}} The suppression of $T_c$ and resistivity with increasing current also warrants further discussion. In previous work it was attributed to the chiral orbital current (COC) state, which is exceedingly sensitive to application of direct current $I$ \cite{zhangControl2022, zhangCurrentsensitive2024}. However, a recent study found that $T_c$ and resistivity was weakly affected by increasing pause current (duration: 0.5 ms) which limits the temperature rise of the sample, indicating that Joule heating effects may contribute to the observed current control of resistivity behaviors \cite{zhangMagneticTransitionInduced2024}. In our study, by establishing thermal equilibrium equation, we can systematically reproduce the decrease of $T_c$ and the suppression of resistivity below $T_c$ with increasing direct current. 

Due to the poor thermal conductivity of \MST, Joule heat accumulates within the sample, causing the sample temperature to be higher than the environmental temperature. The sample temperature and the measured environmental temperature are denoted as $T_{\rm samp}$ and $T_{\rm env}$, respectively, and the temperature difference between them is given by $\Delta T=T_{\rm samp}\!-\!T_{\rm env}$. Per unit time, the Joule heat generated by the sample and the heat dissipated by the sample to the environment are expressed as $Q_J$ and $Q_s$, respectively:
\begin{gather}
    Q_J=\kappa_1 I^2\rho(T_{\rm samp}),\quad Q_s=\kappa_2\Delta T,
\end{gather}
where $\kappa_1$ and $\kappa_2$ are constant coefficients. When thermal equilibrium is established, $Q_J$ equals $Q_s$ and the following equation is obtained:
\begin{gather}
    \frac{\Delta T}{\xi I^2}=\rho(T_{\rm env}+\Delta T),\label{eqn:Tcsolve}
\end{gather}
where $\xi=\kappa_1/\kappa_2$. For each measured temperature $T_{\rm env}$, the resistivity we actually observe is the value at temperature $T_{\rm env}\!+\!\Delta T$. By numerically solving for the $\Delta T$ value at each temperature $T_{\rm env}$, we obtained the current-dependent resistivity at zero field, as shown in Fig. \ref{fig:peakshift}(a), with $\xi$ set to $5\times 10^3$ K/($\rm A^2$$\Omega$m). As the applied direct current $I$ increases from 1 mA to 10 mA, the resistivity peak (or $T_c$) gradually shifts toward lower temperatures, and meanwhile, the resistivity is suppressed below $T_c$.  Notably, as $I$ increases to 10 mA, the resistivity exhibits a rapid jump near $T_c$, similar to the ``first-order transition" in resistivity mentioned in Ref. \cite{zhangControl2022}.

To explain the shift of $T_c$, and the sharp rise in resistivity near $T_c$ with sufficiently large $I$, we define the left-hand side of Eq. (\ref{eqn:Tcsolve}) as a linear function of $\Delta T$, denoted by $f_I(\Delta T)$, and the right-hand side as $\rho'_{T_{\rm env}}(\Delta T)$ which shifts the resistivity $\rho(T)$ to the left by $T_{\rm env}$. $\Delta T$ is then the x-coordinate value of the intersection point between $f_I(\Delta T)$ and $\rho'_{T_{\rm env}}(\Delta T)$.  The slope of  $f_I(\Delta T)$ depends on the current $I$, the larger the current, the smaller the slope. We select two proximate environmental temperatures below $T_c$, denoted as $T_1$ and $T_2$, where $T_2=T_1+ 0.3$ K, and define the x-coordinate values of the intersection points of $f_I(\Delta T)$ with $\rho'_{T_1}(\Delta T)$ and $\rho'_{T_2}(\Delta T)$ as $\Delta T_1$ and $\Delta T_2$, respectively. When a small current $I_1$ is applied, $f_{I_1}$ exhibits a deeper slope, and both the $\Delta T_1$ and $\Delta T_2$ are small, as shown in Fig. \ref{fig:peakshift}(b). This implies a small discrepancy between the measured temperature $T_{\rm env}$ and the sample temperature $T_{\rm samp}$, such that the resistivity peak is observed at a temperature slightly lower than the true $T_c$, indicating a minor shift of $T_c$. Additionally, $\Delta T_1\approx \Delta T_2$ and $\rho'_{T_1}(\Delta T) \approx \rho'_{T_2}(\Delta T)$, suggest that the observed resistivity changes smoothly from $T_1$ to $T_2$. When a larger current $I_2$ is applied, $f_{I_2}$ exhibits a gentler slope. As shown in Fig. \ref{fig:peakshift}(b), $\Delta T_1$ and $\Delta T_2$ become large now, indicating that the resistivity peak can now be observed at a temperature significantly lower than the true $T_c$, resulting in a substantial shift of $T_c$, as shown in Fig. \ref{fig:peakshift}(a). $\Delta T_1$ and $\Delta T_2$ now differ significantly, and the corresponding resistivity values, $\rho'_{T_1}(\Delta T)$ and $\rho'_{T_2}(\Delta T)$, also exhibit a significant disparity. As the measured temperature increases slightly from $T_1$ to $T_2$, the observed resistivity surges from $\rho'_{T_1}(\Delta T)$ to $\rho'_{T_2}(\Delta T)$, thereby elucidating the ``first-order transition" in resistivity at $T_c$ induced by a relatively large current.

In summary, we propose a framework elucidating CMR in magnetic semiconductors as well as the concomitant complex resistivity behaviors, resolving long-standing controversies. By integrating a thermally activated two-carrier model with magnetization-dependent band-gap modulation, we quantitatively reproduce the CMR, anomalous resistivity humps, and their field/temperature-dependent shifts in \MST. Key to this framework is the dynamic modulation of the band gap by magnetic field and temperature, leading to the complex interplay between thermal excitation of electron-hole pairs and magnetization-induced gap variation. Crucially, we demonstrate that Joule heating without considering COC state, results in rather considerable suppression in $T_c$ and the first-order-like transition in resistivity, validated via thermal equilibrium analysis. This work establishes a universal methodology for $\rho(B, T)$ calculations in magnetic semiconductors, providing a comprehensive perspective for unifying all complex phenomena into a single explanatory paradigm.

\textit{Note added:} During the finalization of this manuscript, we became aware of a recent study by Fang \textit{et.al.} [arXiv:2502.11048] that independently arrives at a similar result regarding the pivotal role of Joule heating effects in the first-order-like transition in DC $V$-$I$ measurements \cite{fangElectrothermal2025}. 

This work was supported by the National Natural Science Foundation of China (Grant No.12274436, 11925408, 11921004), the National Key R\&D Program of China (Grant No. 2023YFA1607400, 2022YFA1403800), the Science Center of the National Natural Science Foundation of China (Grant No. 12188101), and  H.W. acknowledge support from the Informatization Plan of the Chinese Academy of Sciences (CASWX2021SF-0102).

\bibliography{main}

\begin{thebibliography}{52}%
\makeatletter
\providecommand \@ifxundefined [1]{%
 \@ifx{#1\undefined}
}%
\providecommand \@ifnum [1]{%
 \ifnum #1\expandafter \@firstoftwo
 \else \expandafter \@secondoftwo
 \fi
}%
\providecommand \@ifx [1]{%
 \ifx #1\expandafter \@firstoftwo
 \else \expandafter \@secondoftwo
 \fi
}%
\providecommand \natexlab [1]{#1}%
\providecommand \enquote  [1]{``#1''}%
\providecommand \bibnamefont  [1]{#1}%
\providecommand \bibfnamefont [1]{#1}%
\providecommand \citenamefont [1]{#1}%
\providecommand \href@noop [0]{\@secondoftwo}%
\providecommand \href [0]{\begingroup \@sanitize@url \@href}%
\providecommand \@href[1]{\@@startlink{#1}\@@href}%
\providecommand \@@href[1]{\endgroup#1\@@endlink}%
\providecommand \@sanitize@url [0]{\catcode `\\12\catcode `\$12\catcode `\&12\catcode `\#12\catcode `\^12\catcode `\_12\catcode `\%12\relax}%
\providecommand \@@startlink[1]{}%
\providecommand \@@endlink[0]{}%
\providecommand \url  [0]{\begingroup\@sanitize@url \@url }%
\providecommand \@url [1]{\endgroup\@href {#1}{\urlprefix }}%
\providecommand \urlprefix  [0]{URL }%
\providecommand \Eprint [0]{\href }%
\providecommand \doibase [0]{http://dx.doi.org/}%
\providecommand \selectlanguage [0]{\@gobble}%
\providecommand \bibinfo  [0]{\@secondoftwo}%
\providecommand \bibfield  [0]{\@secondoftwo}%
\providecommand \translation [1]{[#1]}%
\providecommand \BibitemOpen [0]{}%
\providecommand \bibitemStop [0]{}%
\providecommand \bibitemNoStop [0]{.\EOS\space}%
\providecommand \EOS [0]{\spacefactor3000\relax}%
\providecommand \BibitemShut  [1]{\csname bibitem#1\endcsname}%
\let\auto@bib@innerbib\@empty
%</preamble>
\bibitem [{\citenamefont {Jin}\ \emph {et~al.}(1994{\natexlab{a}})\citenamefont {Jin}, \citenamefont {Tiefel}, \citenamefont {McCormack}, \citenamefont {Fastnacht}, \citenamefont {Ramesh},\ and\ \citenamefont {Chen}}]{jinThousandfold1994}%
  \BibitemOpen
  \bibfield  {author} {\bibinfo {author} {\bibfnamefont {S.}~\bibnamefont {Jin}}, \bibinfo {author} {\bibfnamefont {T.~H.}\ \bibnamefont {Tiefel}}, \bibinfo {author} {\bibfnamefont {M.}~\bibnamefont {McCormack}}, \bibinfo {author} {\bibfnamefont {R.~A.}\ \bibnamefont {Fastnacht}}, \bibinfo {author} {\bibfnamefont {R.}~\bibnamefont {Ramesh}}, \ and\ \bibinfo {author} {\bibfnamefont {L.~H.}\ \bibnamefont {Chen}},\ }\bibfield  {title} {\enquote {\bibinfo {title} {Thousandfold change in resistivity in magnetoresistive {La-Ca-Mn-O Films}},}\ }\href {https://www.science.org/doi/10.1126/science.264.5157.413} {\bibfield  {journal} {\bibinfo  {journal} {Science}\ }\textbf {\bibinfo {volume} {264}},\ \bibinfo {pages} {413--415} (\bibinfo {year} {1994}{\natexlab{a}})}\BibitemShut {NoStop}%
\bibitem [{\citenamefont {Jin}\ \emph {et~al.}(1994{\natexlab{b}})\citenamefont {Jin}, \citenamefont {McCormack}, \citenamefont {Tiefel},\ and\ \citenamefont {Ramesh}}]{jinColossal1994}%
  \BibitemOpen
  \bibfield  {author} {\bibinfo {author} {\bibfnamefont {S.}~\bibnamefont {Jin}}, \bibinfo {author} {\bibfnamefont {M.}~\bibnamefont {McCormack}}, \bibinfo {author} {\bibfnamefont {T.~H.}\ \bibnamefont {Tiefel}}, \ and\ \bibinfo {author} {\bibfnamefont {R.}~\bibnamefont {Ramesh}},\ }\bibfield  {title} {\enquote {\bibinfo {title} {Colossal magnetoresistance in {La-Ca-Mn-O} ferromagnetic thin films (invited)},}\ }\href {https://doi.org/10.1063/1.358119} {\bibfield  {journal} {\bibinfo  {journal} {Journal of Applied Physics}\ }\textbf {\bibinfo {volume} {76}},\ \bibinfo {pages} {6929--6933} (\bibinfo {year} {1994}{\natexlab{b}})}\BibitemShut {NoStop}%
\bibitem [{\citenamefont {Tokura}\ \emph {et~al.}(1994)\citenamefont {Tokura}, \citenamefont {Urushibara}, \citenamefont {Moritomo}, \citenamefont {Arima}, \citenamefont {Asamitsu}, \citenamefont {Kido},\ and\ \citenamefont {Furukawa}}]{tokuraGiant1994}%
  \BibitemOpen
  \bibfield  {author} {\bibinfo {author} {\bibfnamefont {Yoshinori}\ \bibnamefont {Tokura}}, \bibinfo {author} {\bibfnamefont {Akira}\ \bibnamefont {Urushibara}}, \bibinfo {author} {\bibfnamefont {Yutaka}\ \bibnamefont {Moritomo}}, \bibinfo {author} {\bibfnamefont {Takahisa}\ \bibnamefont {Arima}}, \bibinfo {author} {\bibfnamefont {Atsushi}\ \bibnamefont {Asamitsu}}, \bibinfo {author} {\bibfnamefont {Giyu}\ \bibnamefont {Kido}}, \ and\ \bibinfo {author} {\bibfnamefont {Nobuo}\ \bibnamefont {Furukawa}},\ }\bibfield  {title} {\enquote {\bibinfo {title} {Giant magnetotransport phenomena in filling-controlled kondo lattice system: {La$_{1-x}$Sr$_x$MnO$_3$}},}\ }\href {https://journals.jps.jp/doi/10.1143/JPSJ.63.3931} {\bibfield  {journal} {\bibinfo  {journal} {Journal of the Physical Society of Japan}\ }\textbf {\bibinfo {volume} {63}},\ \bibinfo {pages} {3931--3935} (\bibinfo {year} {1994})}\BibitemShut {NoStop}%
\bibitem [{\citenamefont {Asamitsu}\ \emph {et~al.}(1995)\citenamefont {Asamitsu}, \citenamefont {Moritomo}, \citenamefont {Tomioka}, \citenamefont {Arima},\ and\ \citenamefont {Tokura}}]{asamitsustructural1995}%
  \BibitemOpen
  \bibfield  {author} {\bibinfo {author} {\bibfnamefont {A.}~\bibnamefont {Asamitsu}}, \bibinfo {author} {\bibfnamefont {Y.}~\bibnamefont {Moritomo}}, \bibinfo {author} {\bibfnamefont {Y.}~\bibnamefont {Tomioka}}, \bibinfo {author} {\bibfnamefont {T.}~\bibnamefont {Arima}}, \ and\ \bibinfo {author} {\bibfnamefont {Y.}~\bibnamefont {Tokura}},\ }\bibfield  {title} {\enquote {\bibinfo {title} {A structural phase transition induced by an external magnetic field},}\ }\href {https://www.nature.com/articles/373407a0} {\bibfield  {journal} {\bibinfo  {journal} {Nature}\ }\textbf {\bibinfo {volume} {373}},\ \bibinfo {pages} {407--409} (\bibinfo {year} {1995})}\BibitemShut {NoStop}%
\bibitem [{\citenamefont {Urushibara}\ \emph {et~al.}(1995)\citenamefont {Urushibara}, \citenamefont {Moritomo}, \citenamefont {Arima}, \citenamefont {Asamitsu}, \citenamefont {Kido},\ and\ \citenamefont {Tokura}}]{urushibaraInsulatormetal1995}%
  \BibitemOpen
  \bibfield  {author} {\bibinfo {author} {\bibfnamefont {A.}~\bibnamefont {Urushibara}}, \bibinfo {author} {\bibfnamefont {Y.}~\bibnamefont {Moritomo}}, \bibinfo {author} {\bibfnamefont {T.}~\bibnamefont {Arima}}, \bibinfo {author} {\bibfnamefont {A.}~\bibnamefont {Asamitsu}}, \bibinfo {author} {\bibfnamefont {G.}~\bibnamefont {Kido}}, \ and\ \bibinfo {author} {\bibfnamefont {Y.}~\bibnamefont {Tokura}},\ }\bibfield  {title} {\enquote {\bibinfo {title} {Insulator-metal transition and giant magnetoresistance in {La$_{1-x}$Sr$_x$MnO$_3$}},}\ }\href {https://link.aps.org/doi/10.1103/PhysRevB.51.14103} {\bibfield  {journal} {\bibinfo  {journal} {Physical Review B}\ }\textbf {\bibinfo {volume} {51}},\ \bibinfo {pages} {14103--14109} (\bibinfo {year} {1995})}\BibitemShut {NoStop}%
\bibitem [{\citenamefont {R{\"o}der}\ \emph {et~al.}(1996)\citenamefont {R{\"o}der}, \citenamefont {Zang},\ and\ \citenamefont {Bishop}}]{roderLattice1996}%
  \BibitemOpen
  \bibfield  {author} {\bibinfo {author} {\bibfnamefont {H.}~\bibnamefont {R{\"o}der}}, \bibinfo {author} {\bibfnamefont {Jun}\ \bibnamefont {Zang}}, \ and\ \bibinfo {author} {\bibfnamefont {A.~R.}\ \bibnamefont {Bishop}},\ }\bibfield  {title} {\enquote {\bibinfo {title} {Lattice effects in the colossal-magnetoresistance manganites},}\ }\href {https://link.aps.org/doi/10.1103/PhysRevLett.76.1356} {\bibfield  {journal} {\bibinfo  {journal} {Physical Review Letters}\ }\textbf {\bibinfo {volume} {76}},\ \bibinfo {pages} {1356--1359} (\bibinfo {year} {1996})}\BibitemShut {NoStop}%
\bibitem [{\citenamefont {Ramirez}(1997)}]{ramirezColossal1997}%
  \BibitemOpen
  \bibfield  {author} {\bibinfo {author} {\bibfnamefont {A.~P.}\ \bibnamefont {Ramirez}},\ }\bibfield  {title} {\enquote {\bibinfo {title} {Colossal magnetoresistance},}\ }\href {https://dx.doi.org/10.1088/0953-8984/9/39/005} {\bibfield  {journal} {\bibinfo  {journal} {Journal of Physics: Condensed Matter}\ }\textbf {\bibinfo {volume} {9}},\ \bibinfo {pages} {8171} (\bibinfo {year} {1997})}\BibitemShut {NoStop}%
\bibitem [{\citenamefont {Zener}(1951)}]{zenerInteraction1951}%
  \BibitemOpen
  \bibfield  {author} {\bibinfo {author} {\bibfnamefont {Clarence}\ \bibnamefont {Zener}},\ }\bibfield  {title} {\enquote {\bibinfo {title} {Interaction between the $d$-shells in the transition metals. ii. ferromagnetic compounds of manganese with perovskite structure},}\ }\href {https://link.aps.org/doi/10.1103/PhysRev.82.403} {\bibfield  {journal} {\bibinfo  {journal} {Physical Review}\ }\textbf {\bibinfo {volume} {82}},\ \bibinfo {pages} {403--405} (\bibinfo {year} {1951})}\BibitemShut {NoStop}%
\bibitem [{\citenamefont {Anderson}\ and\ \citenamefont {Hasegawa}(1955)}]{andersonConsiderations1955}%
  \BibitemOpen
  \bibfield  {author} {\bibinfo {author} {\bibfnamefont {P.~W.}\ \bibnamefont {Anderson}}\ and\ \bibinfo {author} {\bibfnamefont {H.}~\bibnamefont {Hasegawa}},\ }\bibfield  {title} {\enquote {\bibinfo {title} {Considerations on double exchange},}\ }\href {https://link.aps.org/doi/10.1103/PhysRev.100.675} {\bibfield  {journal} {\bibinfo  {journal} {Physical Review}\ }\textbf {\bibinfo {volume} {100}},\ \bibinfo {pages} {675--681} (\bibinfo {year} {1955})}\BibitemShut {NoStop}%
\bibitem [{\citenamefont {{de Gennes}}(1960)}]{degennesEffects1960}%
  \BibitemOpen
  \bibfield  {author} {\bibinfo {author} {\bibfnamefont {P.~G.}\ \bibnamefont {{de Gennes}}},\ }\bibfield  {title} {\enquote {\bibinfo {title} {Effects of double exchange in magnetic crystals},}\ }\href {https://link.aps.org/doi/10.1103/PhysRev.118.141} {\bibfield  {journal} {\bibinfo  {journal} {Physical Review}\ }\textbf {\bibinfo {volume} {118}},\ \bibinfo {pages} {141--154} (\bibinfo {year} {1960})}\BibitemShut {NoStop}%
\bibitem [{\citenamefont {Millis}\ \emph {et~al.}(1995)\citenamefont {Millis}, \citenamefont {Littlewood},\ and\ \citenamefont {Shraiman}}]{millisDouble1995}%
  \BibitemOpen
  \bibfield  {author} {\bibinfo {author} {\bibfnamefont {A.~J.}\ \bibnamefont {Millis}}, \bibinfo {author} {\bibfnamefont {P.~B.}\ \bibnamefont {Littlewood}}, \ and\ \bibinfo {author} {\bibfnamefont {B.~I.}\ \bibnamefont {Shraiman}},\ }\bibfield  {title} {\enquote {\bibinfo {title} {Double exchange alone does not explain the resistivity of {La$_{1-x}$Sr$_x$MnO$_3$}},}\ }\href {https://link.aps.org/doi/10.1103/PhysRevLett.74.5144} {\bibfield  {journal} {\bibinfo  {journal} {Physical Review Letters}\ }\textbf {\bibinfo {volume} {74}},\ \bibinfo {pages} {5144--5147} (\bibinfo {year} {1995})}\BibitemShut {NoStop}%
\bibitem [{\citenamefont {Millis}\ \emph {et~al.}(1996)\citenamefont {Millis}, \citenamefont {Shraiman},\ and\ \citenamefont {Mueller}}]{millisDynamic1996}%
  \BibitemOpen
  \bibfield  {author} {\bibinfo {author} {\bibfnamefont {A.~J.}\ \bibnamefont {Millis}}, \bibinfo {author} {\bibfnamefont {Boris~I.}\ \bibnamefont {Shraiman}}, \ and\ \bibinfo {author} {\bibfnamefont {R.}~\bibnamefont {Mueller}},\ }\bibfield  {title} {\enquote {\bibinfo {title} {Dynamic jahn-teller effect and colossal magnetoresistance in {La$_{1-x}$Sr$_x$MnO$_3$}},}\ }\href {https://link.aps.org/doi/10.1103/PhysRevLett.77.175} {\bibfield  {journal} {\bibinfo  {journal} {Physical Review Letters}\ }\textbf {\bibinfo {volume} {77}},\ \bibinfo {pages} {175--178} (\bibinfo {year} {1996})}\BibitemShut {NoStop}%
\bibitem [{\citenamefont {Salamon}\ and\ \citenamefont {Jaime}(2001)}]{salamonphysics2001}%
  \BibitemOpen
  \bibfield  {author} {\bibinfo {author} {\bibfnamefont {Myron~B.}\ \bibnamefont {Salamon}}\ and\ \bibinfo {author} {\bibfnamefont {Marcelo}\ \bibnamefont {Jaime}},\ }\bibfield  {title} {\enquote {\bibinfo {title} {The physics of manganites: Structure and transport},}\ }\href {https://link.aps.org/doi/10.1103/RevModPhys.73.583} {\bibfield  {journal} {\bibinfo  {journal} {Reviews of Modern Physics}\ }\textbf {\bibinfo {volume} {73}},\ \bibinfo {pages} {583--628} (\bibinfo {year} {2001})}\BibitemShut {NoStop}%
\bibitem [{\citenamefont {Shapira}\ and\ \citenamefont {Reed}(1972)}]{shapiraResistivity1972}%
  \BibitemOpen
  \bibfield  {author} {\bibinfo {author} {\bibfnamefont {Y.}~\bibnamefont {Shapira}}\ and\ \bibinfo {author} {\bibfnamefont {T.~B.}\ \bibnamefont {Reed}},\ }\bibfield  {title} {\enquote {\bibinfo {title} {Resistivity and hall effect of {EuS} in fields up to 140 {kOe}},}\ }\href {https://link.aps.org/doi/10.1103/PhysRevB.5.4877} {\bibfield  {journal} {\bibinfo  {journal} {Physical Review B}\ }\textbf {\bibinfo {volume} {5}},\ \bibinfo {pages} {4877--4890} (\bibinfo {year} {1972})}\BibitemShut {NoStop}%
\bibitem [{\citenamefont {Oliver}\ \emph {et~al.}(1972)\citenamefont {Oliver}, \citenamefont {Dimmock}, \citenamefont {McWhorter},\ and\ \citenamefont {Reed}}]{oliverConductivity1972}%
  \BibitemOpen
  \bibfield  {author} {\bibinfo {author} {\bibfnamefont {M.~R.}\ \bibnamefont {Oliver}}, \bibinfo {author} {\bibfnamefont {J.~O.}\ \bibnamefont {Dimmock}}, \bibinfo {author} {\bibfnamefont {A.~L.}\ \bibnamefont {McWhorter}}, \ and\ \bibinfo {author} {\bibfnamefont {T.~B.}\ \bibnamefont {Reed}},\ }\bibfield  {title} {\enquote {\bibinfo {title} {Conductivity studies in europium oxide},}\ }\href {https://link.aps.org/doi/10.1103/PhysRevB.5.1078} {\bibfield  {journal} {\bibinfo  {journal} {Physical Review B}\ }\textbf {\bibinfo {volume} {5}},\ \bibinfo {pages} {1078--1098} (\bibinfo {year} {1972})}\BibitemShut {NoStop}%
\bibitem [{\citenamefont {Shapira}\ \emph {et~al.}(1973)\citenamefont {Shapira}, \citenamefont {Foner},\ and\ \citenamefont {Reed}}]{shapiraEuO1973}%
  \BibitemOpen
  \bibfield  {author} {\bibinfo {author} {\bibfnamefont {Y.}~\bibnamefont {Shapira}}, \bibinfo {author} {\bibfnamefont {S.}~\bibnamefont {Foner}}, \ and\ \bibinfo {author} {\bibfnamefont {T.~B.}\ \bibnamefont {Reed}},\ }\bibfield  {title} {\enquote {\bibinfo {title} {{EuO}. i. resistivity and hall effect in fields up to 150 {kOe}},}\ }\href {https://link.aps.org/doi/10.1103/PhysRevB.8.2299} {\bibfield  {journal} {\bibinfo  {journal} {Physical Review B}\ }\textbf {\bibinfo {volume} {8}},\ \bibinfo {pages} {2299--2315} (\bibinfo {year} {1973})}\BibitemShut {NoStop}%
\bibitem [{\citenamefont {Konno}\ \emph {et~al.}(1998)\citenamefont {Konno}, \citenamefont {Wakoh}, \citenamefont {Sumiyama},\ and\ \citenamefont {Suzuki}}]{konnoElectrical1998}%
  \BibitemOpen
  \bibfield  {author} {\bibinfo {author} {\bibfnamefont {Toyohiko~J.}\ \bibnamefont {Konno}}, \bibinfo {author} {\bibfnamefont {Kimio}\ \bibnamefont {Wakoh}}, \bibinfo {author} {\bibfnamefont {Kenji}\ \bibnamefont {Sumiyama}}, \ and\ \bibinfo {author} {\bibfnamefont {Kenji}\ \bibnamefont {Suzuki}},\ }\bibfield  {title} {\enquote {\bibinfo {title} {Electrical resistivity of {Eu}-rich {EuO} thin films},}\ }\href {https://iopscience.iop.org/article/10.1143/JJAP.37.L787/meta} {\bibfield  {journal} {\bibinfo  {journal} {Japanese Journal of Applied Physics}\ }\textbf {\bibinfo {volume} {37}},\ \bibinfo {pages} {L787} (\bibinfo {year} {1998})}\BibitemShut {NoStop}%
\bibitem [{\citenamefont {Sun}\ \emph {et~al.}(2010)\citenamefont {Sun}, \citenamefont {Huang}, \citenamefont {Lin}, \citenamefont {Her}, \citenamefont {Ho}, \citenamefont {Lin}, \citenamefont {Berger},\ and\ \citenamefont {Yang}}]{sunColossal2010}%
  \BibitemOpen
  \bibfield  {author} {\bibinfo {author} {\bibfnamefont {C.~P.}\ \bibnamefont {Sun}}, \bibinfo {author} {\bibfnamefont {C.~L.}\ \bibnamefont {Huang}}, \bibinfo {author} {\bibfnamefont {C.~C.}\ \bibnamefont {Lin}}, \bibinfo {author} {\bibfnamefont {J.~L.}\ \bibnamefont {Her}}, \bibinfo {author} {\bibfnamefont {C.~J.}\ \bibnamefont {Ho}}, \bibinfo {author} {\bibfnamefont {J.-Y.}\ \bibnamefont {Lin}}, \bibinfo {author} {\bibfnamefont {H.}~\bibnamefont {Berger}}, \ and\ \bibinfo {author} {\bibfnamefont {H.~D.}\ \bibnamefont {Yang}},\ }\bibfield  {title} {\enquote {\bibinfo {title} {Colossal electroresistance and colossal magnetoresistance in spinel multiferroic {CdCr$_2$S$_4$}},}\ }\href {https://doi.org/10.1063/1.3368123} {\bibfield  {journal} {\bibinfo  {journal} {Applied Physics Letters}\ }\textbf {\bibinfo {volume} {96}},\ \bibinfo {pages} {122109} (\bibinfo {year} {2010})}\BibitemShut {NoStop}%
\bibitem [{\citenamefont {Lin}\ \emph {et~al.}(2016)\citenamefont {Lin}, \citenamefont {Yi}, \citenamefont {Shi}, \citenamefont {Zhang}, \citenamefont {Zhang}, \citenamefont {M{\"u}ller},\ and\ \citenamefont {Li}}]{linSpin2016a}%
  \BibitemOpen
  \bibfield  {author} {\bibinfo {author} {\bibfnamefont {Chaojing}\ \bibnamefont {Lin}}, \bibinfo {author} {\bibfnamefont {Changjiang}\ \bibnamefont {Yi}}, \bibinfo {author} {\bibfnamefont {Youguo}\ \bibnamefont {Shi}}, \bibinfo {author} {\bibfnamefont {Lei}\ \bibnamefont {Zhang}}, \bibinfo {author} {\bibfnamefont {Guangming}\ \bibnamefont {Zhang}}, \bibinfo {author} {\bibfnamefont {Jens}\ \bibnamefont {M{\"u}ller}}, \ and\ \bibinfo {author} {\bibfnamefont {Yongqing}\ \bibnamefont {Li}},\ }\bibfield  {title} {\enquote {\bibinfo {title} {Spin correlations and colossal magnetoresistance in {HgCr$_2$Se$_4$ }},}\ }\href {https://link.aps.org/doi/10.1103/PhysRevB.94.224404} {\bibfield  {journal} {\bibinfo  {journal} {Physical Review B}\ }\textbf {\bibinfo {volume} {94}},\ \bibinfo {pages} {224404} (\bibinfo {year} {2016})}\BibitemShut {NoStop}%
\bibitem [{\citenamefont {Nagaev}(1999)}]{nagaevMagnetoimpurity1999}%
  \BibitemOpen
  \bibfield  {author} {\bibinfo {author} {\bibfnamefont {E.~L.}\ \bibnamefont {Nagaev}},\ }\bibfield  {title} {\enquote {\bibinfo {title} {Magnetoimpurity theory of manganites and other colossal magnetoresistance materials},}\ }\href {http://www.publish.csiro.au/?paper=P98074} {\bibfield  {journal} {\bibinfo  {journal} {Australian Journal of Physics}\ }\textbf {\bibinfo {volume} {52}},\ \bibinfo {pages} {305} (\bibinfo {year} {1999})}\BibitemShut {NoStop}%
\bibitem [{\citenamefont {Nagaev}(2001)}]{nagaevColossalmagnetoresistance2001}%
  \BibitemOpen
  \bibfield  {author} {\bibinfo {author} {\bibfnamefont {E.~L.}\ \bibnamefont {Nagaev}},\ }\bibfield  {title} {\enquote {\bibinfo {title} {Colossal-magnetoresistance materials: Manganites and conventional ferromagnetic semiconductors},}\ }\href {https://www.sciencedirect.com/science/article/pii/S0370157300001113} {\bibfield  {journal} {\bibinfo  {journal} {Physics Reports}\ }\textbf {\bibinfo {volume} {346}},\ \bibinfo {pages} {387--531} (\bibinfo {year} {2001})}\BibitemShut {NoStop}%
\bibitem [{\citenamefont {Majumdar}\ and\ \citenamefont {Littlewood}(1998)}]{majumdarMagnetoresistance1998}%
  \BibitemOpen
  \bibfield  {author} {\bibinfo {author} {\bibfnamefont {Pinaki}\ \bibnamefont {Majumdar}}\ and\ \bibinfo {author} {\bibfnamefont {Peter}\ \bibnamefont {Littlewood}},\ }\bibfield  {title} {\enquote {\bibinfo {title} {Magnetoresistance in mn pyrochlore: Electrical transport in a low carrier density ferromagnet},}\ }\href {https://link.aps.org/doi/10.1103/PhysRevLett.81.1314} {\bibfield  {journal} {\bibinfo  {journal} {Physical Review Letters}\ }\textbf {\bibinfo {volume} {81}},\ \bibinfo {pages} {1314--1317} (\bibinfo {year} {1998})}\BibitemShut {NoStop}%
\bibitem [{\citenamefont {S{\"u}llow}\ \emph {et~al.}(2000)\citenamefont {S{\"u}llow}, \citenamefont {Prasad}, \citenamefont {Bogdanovich}, \citenamefont {Aronson}, \citenamefont {Sarrao},\ and\ \citenamefont {Fisk}}]{sullowMagnetotransport2000}%
  \BibitemOpen
  \bibfield  {author} {\bibinfo {author} {\bibfnamefont {S.}~\bibnamefont {S{\"u}llow}}, \bibinfo {author} {\bibfnamefont {I.}~\bibnamefont {Prasad}}, \bibinfo {author} {\bibfnamefont {S.}~\bibnamefont {Bogdanovich}}, \bibinfo {author} {\bibfnamefont {M.~C.}\ \bibnamefont {Aronson}}, \bibinfo {author} {\bibfnamefont {J.~L.}\ \bibnamefont {Sarrao}}, \ and\ \bibinfo {author} {\bibfnamefont {Z.}~\bibnamefont {Fisk}},\ }\bibfield  {title} {\enquote {\bibinfo {title} {Magnetotransport in the low carrier density ferromagnet {EuB$_6$}},}\ }\href {https://doi.org/10.1063/1.372460} {\bibfield  {journal} {\bibinfo  {journal} {Journal of Applied Physics}\ }\textbf {\bibinfo {volume} {87}},\ \bibinfo {pages} {5591--5593} (\bibinfo {year} {2000})}\BibitemShut {NoStop}%
\bibitem [{\citenamefont {Yang}\ \emph {et~al.}(2004)\citenamefont {Yang}, \citenamefont {Bao}, \citenamefont {Tan},\ and\ \citenamefont {Zhang}}]{yangMagnetic2004}%
  \BibitemOpen
  \bibfield  {author} {\bibinfo {author} {\bibfnamefont {Zhaorong}\ \bibnamefont {Yang}}, \bibinfo {author} {\bibfnamefont {Xinyu}\ \bibnamefont {Bao}}, \bibinfo {author} {\bibfnamefont {Shun}\ \bibnamefont {Tan}}, \ and\ \bibinfo {author} {\bibfnamefont {Yuheng}\ \bibnamefont {Zhang}},\ }\bibfield  {title} {\enquote {\bibinfo {title} {{Magnetic Polaron Conduction in the Colossal Magnetoresistance Material {Fe$_{1-x}$Cd$_x$Cr$_2$S$_4$}}},}\ }\href {https://link.aps.org/doi/10.1103/PhysRevB.69.144407} {\bibfield  {journal} {\bibinfo  {journal} {Physical Review B}\ }\textbf {\bibinfo {volume} {69}},\ \bibinfo {pages} {144407} (\bibinfo {year} {2004})}\BibitemShut {NoStop}%
\bibitem [{\citenamefont {Pohlit}\ \emph {et~al.}(2018)\citenamefont {Pohlit}, \citenamefont {R{\"o}{\ss}ler}, \citenamefont {Ohno}, \citenamefont {Ohno}, \citenamefont {{von Moln{\'a}r}}, \citenamefont {Fisk}, \citenamefont {M{\"u}ller},\ and\ \citenamefont {Wirth}}]{pohlitEvidence2018}%
  \BibitemOpen
  \bibfield  {author} {\bibinfo {author} {\bibfnamefont {Merlin}\ \bibnamefont {Pohlit}}, \bibinfo {author} {\bibfnamefont {Sahana}\ \bibnamefont {R{\"o}{\ss}ler}}, \bibinfo {author} {\bibfnamefont {Yuzo}\ \bibnamefont {Ohno}}, \bibinfo {author} {\bibfnamefont {Hideo}\ \bibnamefont {Ohno}}, \bibinfo {author} {\bibfnamefont {Stephan}\ \bibnamefont {{von Moln{\'a}r}}}, \bibinfo {author} {\bibfnamefont {Zachary}\ \bibnamefont {Fisk}}, \bibinfo {author} {\bibfnamefont {Jens}\ \bibnamefont {M{\"u}ller}}, \ and\ \bibinfo {author} {\bibfnamefont {Steffen}\ \bibnamefont {Wirth}},\ }\bibfield  {title} {\enquote {\bibinfo {title} {Evidence for ferromagnetic clusters in the colossal-magnetoresistance material {EuB$_6$}},}\ }\href {https://link.aps.org/doi/10.1103/PhysRevLett.120.257201} {\bibfield  {journal} {\bibinfo  {journal} {Physical Review Letters}\ }\textbf {\bibinfo {volume} {120}},\ \bibinfo {pages} {257201} (\bibinfo {year} {2018})}\BibitemShut {NoStop}%
\bibitem [{\citenamefont {Tomioka}\ \emph {et~al.}(1997)\citenamefont {Tomioka}, \citenamefont {Asamitsu}, \citenamefont {Kuwahara}, \citenamefont {Moritomo}, \citenamefont {Kasai}, \citenamefont {Kumai},\ and\ \citenamefont {Tokura}}]{tomiokaMagneticfieldinduced1997}%
  \BibitemOpen
  \bibfield  {author} {\bibinfo {author} {\bibfnamefont {Y.}~\bibnamefont {Tomioka}}, \bibinfo {author} {\bibfnamefont {A.}~\bibnamefont {Asamitsu}}, \bibinfo {author} {\bibfnamefont {H.}~\bibnamefont {Kuwahara}}, \bibinfo {author} {\bibfnamefont {Y.}~\bibnamefont {Moritomo}}, \bibinfo {author} {\bibfnamefont {M.}~\bibnamefont {Kasai}}, \bibinfo {author} {\bibfnamefont {R.}~\bibnamefont {Kumai}}, \ and\ \bibinfo {author} {\bibfnamefont {Y.}~\bibnamefont {Tokura}},\ }\bibfield  {title} {\enquote {\bibinfo {title} {Magnetic-field-induced metal-insulator transition in perovskite-type manganese oxides},}\ }\href {https://www.sciencedirect.com/science/article/pii/S0921452697000136} {\bibfield  {journal} {\bibinfo  {journal} {Physica B: Condensed Matter}\ }\bibinfo {series} {Proceedings of the Yamada Conference XLV, the International Conference on the Physics of Transition Metals},\ \textbf {\bibinfo {volume} {237--238}},\ \bibinfo {pages} {6--10} (\bibinfo {year} {1997})}\BibitemShut {NoStop}%
\bibitem [{\citenamefont {Yunoki}\ and\ \citenamefont {Moreo}(1998)}]{yunokiStatic1998}%
  \BibitemOpen
  \bibfield  {author} {\bibinfo {author} {\bibfnamefont {Seiji}\ \bibnamefont {Yunoki}}\ and\ \bibinfo {author} {\bibfnamefont {Adriana}\ \bibnamefont {Moreo}},\ }\bibfield  {title} {\enquote {\bibinfo {title} {Static and dynamical properties of the ferromagnetic kondo model with direct antiferromagnetic coupling between the localized $t_{2g}$ electrons},}\ }\href {https://link.aps.org/doi/10.1103/PhysRevB.58.6403} {\bibfield  {journal} {\bibinfo  {journal} {Physical Review B}\ }\textbf {\bibinfo {volume} {58}},\ \bibinfo {pages} {6403--6413} (\bibinfo {year} {1998})}\BibitemShut {NoStop}%
\bibitem [{\citenamefont {Alonso}\ \emph {et~al.}(2001)\citenamefont {Alonso}, \citenamefont {Fern{\'a}ndez}, \citenamefont {Guinea}, \citenamefont {Laliena},\ and\ \citenamefont {{Mart{\'i}n-Mayor}}}]{alonsoDiscontinuous2001}%
  \BibitemOpen
  \bibfield  {author} {\bibinfo {author} {\bibfnamefont {J.~L.}\ \bibnamefont {Alonso}}, \bibinfo {author} {\bibfnamefont {L.~A.}\ \bibnamefont {Fern{\'a}ndez}}, \bibinfo {author} {\bibfnamefont {F.}~\bibnamefont {Guinea}}, \bibinfo {author} {\bibfnamefont {V.}~\bibnamefont {Laliena}}, \ and\ \bibinfo {author} {\bibfnamefont {V.}~\bibnamefont {{Mart{\'i}n-Mayor}}},\ }\bibfield  {title} {\enquote {\bibinfo {title} {Discontinuous transitions in double-exchange materials},}\ }\href {https://link.aps.org/doi/10.1103/PhysRevB.63.064416} {\bibfield  {journal} {\bibinfo  {journal} {Physical Review B}\ }\textbf {\bibinfo {volume} {63}},\ \bibinfo {pages} {064416} (\bibinfo {year} {2001})}\BibitemShut {NoStop}%
\bibitem [{\citenamefont {Dagotto}\ \emph {et~al.}(2001)\citenamefont {Dagotto}, \citenamefont {Hotta},\ and\ \citenamefont {Moreo}}]{dagottoColossal2001}%
  \BibitemOpen
  \bibfield  {author} {\bibinfo {author} {\bibfnamefont {Elbio}\ \bibnamefont {Dagotto}}, \bibinfo {author} {\bibfnamefont {Takashi}\ \bibnamefont {Hotta}}, \ and\ \bibinfo {author} {\bibfnamefont {Adriana}\ \bibnamefont {Moreo}},\ }\bibfield  {title} {\enquote {\bibinfo {title} {Colossal magnetoresistant materials: The key role of phase separation},}\ }\href {https://www.sciencedirect.com/science/article/pii/S0370157300001216} {\bibfield  {journal} {\bibinfo  {journal} {Physics Reports}\ }\textbf {\bibinfo {volume} {344}},\ \bibinfo {pages} {1--153} (\bibinfo {year} {2001})}\BibitemShut {NoStop}%
\bibitem [{\citenamefont {Tokura}(2006)}]{tokuraCritical2006}%
  \BibitemOpen
  \bibfield  {author} {\bibinfo {author} {\bibfnamefont {Y}~\bibnamefont {Tokura}},\ }\bibfield  {title} {\enquote {\bibinfo {title} {Critical features of colossal magnetoresistive manganites},}\ }\href {https://dx.doi.org/10.1088/0034-4885/69/3/R06} {\bibfield  {journal} {\bibinfo  {journal} {Reports on Progress in Physics}\ }\textbf {\bibinfo {volume} {69}},\ \bibinfo {pages} {797} (\bibinfo {year} {2006})}\BibitemShut {NoStop}%
\bibitem [{\citenamefont {{\c S}en}\ \emph {et~al.}(2007)\citenamefont {{\c S}en}, \citenamefont {Alvarez},\ and\ \citenamefont {Dagotto}}]{senCompeting2007}%
  \BibitemOpen
  \bibfield  {author} {\bibinfo {author} {\bibfnamefont {Cengiz}\ \bibnamefont {{\c S}en}}, \bibinfo {author} {\bibfnamefont {Gonzalo}\ \bibnamefont {Alvarez}}, \ and\ \bibinfo {author} {\bibfnamefont {Elbio}\ \bibnamefont {Dagotto}},\ }\bibfield  {title} {\enquote {\bibinfo {title} {Competing ferromagnetic and charge-ordered states in models for manganites: The origin of the colossal magnetoresistance effect},}\ }\href {https://link.aps.org/doi/10.1103/PhysRevLett.98.127202} {\bibfield  {journal} {\bibinfo  {journal} {Physical Review Letters}\ }\textbf {\bibinfo {volume} {98}},\ \bibinfo {pages} {127202} (\bibinfo {year} {2007})}\BibitemShut {NoStop}%
\bibitem [{\citenamefont {Weber}\ \emph {et~al.}(2006)\citenamefont {Weber}, \citenamefont {Lunkenheimer}, \citenamefont {Fichtl}, \citenamefont {Hemberger}, \citenamefont {Tsurkan},\ and\ \citenamefont {Loidl}}]{weberColossal2006}%
  \BibitemOpen
  \bibfield  {author} {\bibinfo {author} {\bibfnamefont {S.}~\bibnamefont {Weber}}, \bibinfo {author} {\bibfnamefont {P.}~\bibnamefont {Lunkenheimer}}, \bibinfo {author} {\bibfnamefont {R.}~\bibnamefont {Fichtl}}, \bibinfo {author} {\bibfnamefont {J.}~\bibnamefont {Hemberger}}, \bibinfo {author} {\bibfnamefont {V.}~\bibnamefont {Tsurkan}}, \ and\ \bibinfo {author} {\bibfnamefont {A.}~\bibnamefont {Loidl}},\ }\bibfield  {title} {\enquote {\bibinfo {title} {Colossal magnetocapacitance and colossal magnetoresistance in {HgCr$_2$S$_4$}},}\ }\href {https://doi.org/10.1103/PhysRevLett.96.157202} {\bibfield  {journal} {\bibinfo  {journal} {Physical Review Letters}\ }\textbf {\bibinfo {volume} {96}},\ \bibinfo {pages} {157202} (\bibinfo {year} {2006})}\BibitemShut {NoStop}%
\bibitem [{\citenamefont {Rosa}\ \emph {et~al.}(2020)\citenamefont {Rosa}, \citenamefont {Xu}, \citenamefont {Rahn}, \citenamefont {Souza}, \citenamefont {Kushwaha}, \citenamefont {Veiga}, \citenamefont {Bombardi}, \citenamefont {Thomas}, \citenamefont {Janoschek}, \citenamefont {Bauer}, \citenamefont {Chan}, \citenamefont {Wang}, \citenamefont {Thompson}, \citenamefont {Harrison}, \citenamefont {Pagliuso}, \citenamefont {Bernevig},\ and\ \citenamefont {Ronning}}]{rosaColossal2020}%
  \BibitemOpen
  \bibfield  {author} {\bibinfo {author} {\bibfnamefont {Priscila}\ \bibnamefont {Rosa}}, \bibinfo {author} {\bibfnamefont {Yuanfeng}\ \bibnamefont {Xu}}, \bibinfo {author} {\bibfnamefont {Marein}\ \bibnamefont {Rahn}}, \bibinfo {author} {\bibfnamefont {Jean}\ \bibnamefont {Souza}}, \bibinfo {author} {\bibfnamefont {Satya}\ \bibnamefont {Kushwaha}}, \bibinfo {author} {\bibfnamefont {Larissa}\ \bibnamefont {Veiga}}, \bibinfo {author} {\bibfnamefont {Alessandro}\ \bibnamefont {Bombardi}}, \bibinfo {author} {\bibfnamefont {Sean}\ \bibnamefont {Thomas}}, \bibinfo {author} {\bibfnamefont {Marc}\ \bibnamefont {Janoschek}}, \bibinfo {author} {\bibfnamefont {Eric}\ \bibnamefont {Bauer}}, \bibinfo {author} {\bibfnamefont {Mun}\ \bibnamefont {Chan}}, \bibinfo {author} {\bibfnamefont {Zhijun}\ \bibnamefont {Wang}}, \bibinfo {author} {\bibfnamefont {Joe}\ \bibnamefont {Thompson}}, \bibinfo {author} {\bibfnamefont {Neil}\ \bibnamefont {Harrison}}, \bibinfo {author} {\bibfnamefont {Pascoal}\ \bibnamefont {Pagliuso}},
  \bibinfo {author} {\bibfnamefont {Andrei}\ \bibnamefont {Bernevig}}, \ and\ \bibinfo {author} {\bibfnamefont {Filip}\ \bibnamefont {Ronning}},\ }\bibfield  {title} {\enquote {\bibinfo {title} {Colossal magnetoresistance in a nonsymmorphic antiferromagnetic insulator},}\ }\href {https://www.nature.com/articles/s41535-020-00256-8} {\bibfield  {journal} {\bibinfo  {journal} {npj Quantum Materials}\ }\textbf {\bibinfo {volume} {5}},\ \bibinfo {pages} {1--6} (\bibinfo {year} {2020})}\BibitemShut {NoStop}%
\bibitem [{\citenamefont {Wang}\ \emph {et~al.}(2020)\citenamefont {Wang}, \citenamefont {Chang}, \citenamefont {Zhou}, \citenamefont {Ma}, \citenamefont {Lin}, \citenamefont {Hasan}, \citenamefont {Xu},\ and\ \citenamefont {Jia}}]{wangFieldInduced2020}%
  \BibitemOpen
  \bibfield  {author} {\bibinfo {author} {\bibfnamefont {Guangqiang}\ \bibnamefont {Wang}}, \bibinfo {author} {\bibfnamefont {Guoqing}\ \bibnamefont {Chang}}, \bibinfo {author} {\bibfnamefont {Huibin}\ \bibnamefont {Zhou}}, \bibinfo {author} {\bibfnamefont {Wenlong}\ \bibnamefont {Ma}}, \bibinfo {author} {\bibfnamefont {Hsin}\ \bibnamefont {Lin}}, \bibinfo {author} {\bibfnamefont {M.~Zahid}\ \bibnamefont {Hasan}}, \bibinfo {author} {\bibfnamefont {Su-Yang}\ \bibnamefont {Xu}}, \ and\ \bibinfo {author} {\bibfnamefont {Shuang}\ \bibnamefont {Jia}},\ }\bibfield  {title} {\enquote {\bibinfo {title} {Field-induced metal--insulator transition in {$\beta$}-{EuP$_3$}},}\ }\href {https://dx.doi.org/10.1088/0256-307X/37/10/107501} {\bibfield  {journal} {\bibinfo  {journal} {Chinese Physics Letters}\ }\textbf {\bibinfo {volume} {37}},\ \bibinfo {pages} {107501} (\bibinfo {year} {2020})}\BibitemShut {NoStop}%
\bibitem [{\citenamefont {Ni}\ \emph {et~al.}(2021)\citenamefont {Ni}, \citenamefont {Zhao}, \citenamefont {Zhang}, \citenamefont {Hu}, \citenamefont {Kimchi},\ and\ \citenamefont {Cao}}]{niColossal2021}%
  \BibitemOpen
  \bibfield  {author} {\bibinfo {author} {\bibfnamefont {Yifei}\ \bibnamefont {Ni}}, \bibinfo {author} {\bibfnamefont {Hengdi}\ \bibnamefont {Zhao}}, \bibinfo {author} {\bibfnamefont {Yu}~\bibnamefont {Zhang}}, \bibinfo {author} {\bibfnamefont {Bing}\ \bibnamefont {Hu}}, \bibinfo {author} {\bibfnamefont {Itamar}\ \bibnamefont {Kimchi}}, \ and\ \bibinfo {author} {\bibfnamefont {Gang}\ \bibnamefont {Cao}},\ }\bibfield  {title} {\enquote {\bibinfo {title} {Colossal magnetoresistance via avoiding fully polarized magnetization in the ferrimagnetic insulator {Mn$_3$Si$_2$Te$_6$}},}\ }\href {https://link.aps.org/doi/10.1103/PhysRevB.103.L161105} {\bibfield  {journal} {\bibinfo  {journal} {Physical Review B}\ }\textbf {\bibinfo {volume} {103}},\ \bibinfo {pages} {L161105} (\bibinfo {year} {2021})}\BibitemShut {NoStop}%
\bibitem [{\citenamefont {Yin}\ \emph {et~al.}(2020)\citenamefont {Yin}, \citenamefont {Wu}, \citenamefont {Li}, \citenamefont {Yu}, \citenamefont {Sun}, \citenamefont {Shen}, \citenamefont {Frandsen}, \citenamefont {Yao},\ and\ \citenamefont {Wang}}]{yinLarge2020}%
  \BibitemOpen
  \bibfield  {author} {\bibinfo {author} {\bibfnamefont {Junjie}\ \bibnamefont {Yin}}, \bibinfo {author} {\bibfnamefont {Changwei}\ \bibnamefont {Wu}}, \bibinfo {author} {\bibfnamefont {Lisi}\ \bibnamefont {Li}}, \bibinfo {author} {\bibfnamefont {Jia}\ \bibnamefont {Yu}}, \bibinfo {author} {\bibfnamefont {Hualei}\ \bibnamefont {Sun}}, \bibinfo {author} {\bibfnamefont {Bing}\ \bibnamefont {Shen}}, \bibinfo {author} {\bibfnamefont {Benjamin~A.}\ \bibnamefont {Frandsen}}, \bibinfo {author} {\bibfnamefont {Dao-Xin}\ \bibnamefont {Yao}}, \ and\ \bibinfo {author} {\bibfnamefont {Meng}\ \bibnamefont {Wang}},\ }\bibfield  {title} {\enquote {\bibinfo {title} {Large negative magnetoresistance in the antiferromagnetic rare-earth dichalcogenide {EuTe$_2$}},}\ }\href {https://link.aps.org/doi/10.1103/PhysRevMaterials.4.013405} {\bibfield  {journal} {\bibinfo  {journal} {Physical Review Materials}\ }\textbf {\bibinfo {volume} {4}},\ \bibinfo {pages} {013405} (\bibinfo {year} {2020})}\BibitemShut {NoStop}%
\bibitem [{\citenamefont {Yang}\ \emph {et~al.}(2021)\citenamefont {Yang}, \citenamefont {Liu}, \citenamefont {Liao}, \citenamefont {Si}, \citenamefont {Jiang}, \citenamefont {Liu}, \citenamefont {Guo}, \citenamefont {Yin}, \citenamefont {Wang}, \citenamefont {Sheng}, \citenamefont {Zhao}, \citenamefont {Wang}, \citenamefont {Zhong},\ and\ \citenamefont {Li}}]{yangColossal2021}%
  \BibitemOpen
  \bibfield  {author} {\bibinfo {author} {\bibfnamefont {Huali}\ \bibnamefont {Yang}}, \bibinfo {author} {\bibfnamefont {Qing}\ \bibnamefont {Liu}}, \bibinfo {author} {\bibfnamefont {Zhaoliang}\ \bibnamefont {Liao}}, \bibinfo {author} {\bibfnamefont {Liang}\ \bibnamefont {Si}}, \bibinfo {author} {\bibfnamefont {Peiheng}\ \bibnamefont {Jiang}}, \bibinfo {author} {\bibfnamefont {Xiaolei}\ \bibnamefont {Liu}}, \bibinfo {author} {\bibfnamefont {Yanfeng}\ \bibnamefont {Guo}}, \bibinfo {author} {\bibfnamefont {Junjie}\ \bibnamefont {Yin}}, \bibinfo {author} {\bibfnamefont {Meng}\ \bibnamefont {Wang}}, \bibinfo {author} {\bibfnamefont {Zhigao}\ \bibnamefont {Sheng}}, \bibinfo {author} {\bibfnamefont {Yuxin}\ \bibnamefont {Zhao}}, \bibinfo {author} {\bibfnamefont {Zhiming}\ \bibnamefont {Wang}}, \bibinfo {author} {\bibfnamefont {Zhicheng}\ \bibnamefont {Zhong}}, \ and\ \bibinfo {author} {\bibfnamefont {Run-Wei}\ \bibnamefont {Li}},\ }\bibfield  {title} {\enquote {\bibinfo {title} {Colossal angular magnetoresistance
  in the antiferromagnetic semiconductor {EuTe$_2$}},}\ }\href {https://link.aps.org/doi/10.1103/PhysRevB.104.214419} {\bibfield  {journal} {\bibinfo  {journal} {Physical Review B}\ }\textbf {\bibinfo {volume} {104}},\ \bibinfo {pages} {214419} (\bibinfo {year} {2021})}\BibitemShut {NoStop}%
\bibitem [{\citenamefont {Takeuchi}\ \emph {et~al.}(2024)\citenamefont {Takeuchi}, \citenamefont {Honda}, \citenamefont {Aoki}, \citenamefont {Haga}, \citenamefont {Kida}, \citenamefont {Narumi}, \citenamefont {Hagiwara}, \citenamefont {Kindo}, \citenamefont {Karube}, \citenamefont {Harima},\ and\ \citenamefont {{\=O}nuki}}]{takeuchiFieldinduced2024}%
  \BibitemOpen
  \bibfield  {author} {\bibinfo {author} {\bibfnamefont {Tetsuya}\ \bibnamefont {Takeuchi}}, \bibinfo {author} {\bibfnamefont {Fuminori}\ \bibnamefont {Honda}}, \bibinfo {author} {\bibfnamefont {Dai}\ \bibnamefont {Aoki}}, \bibinfo {author} {\bibfnamefont {Yoshinori}\ \bibnamefont {Haga}}, \bibinfo {author} {\bibfnamefont {Takanori}\ \bibnamefont {Kida}}, \bibinfo {author} {\bibfnamefont {Yasuo}\ \bibnamefont {Narumi}}, \bibinfo {author} {\bibfnamefont {Masayuki}\ \bibnamefont {Hagiwara}}, \bibinfo {author} {\bibfnamefont {Koichi}\ \bibnamefont {Kindo}}, \bibinfo {author} {\bibfnamefont {Kosuke}\ \bibnamefont {Karube}}, \bibinfo {author} {\bibfnamefont {Hisatomo}\ \bibnamefont {Harima}}, \ and\ \bibinfo {author} {\bibfnamefont {Yoshichika}\ \bibnamefont {{\=O}nuki}},\ }\bibfield  {title} {\enquote {\bibinfo {title} {Field-induced insulator-metal transition in {EuTe$_2$}},}\ }\href {https://journals.jps.jp/doi/full/10.7566/JPSJ.93.044708} {\bibfield  {journal} {\bibinfo  {journal} {Journal of the Physical
  Society of Japan}\ }\textbf {\bibinfo {volume} {93}},\ \bibinfo {pages} {044708} (\bibinfo {year} {2024})}\BibitemShut {NoStop}%
\bibitem [{\citenamefont {Dong}\ \emph {et~al.}(2024)\citenamefont {Dong}, \citenamefont {Yang}, \citenamefont {Liu}, \citenamefont {Wang}, \citenamefont {Liu}, \citenamefont {Shi}, \citenamefont {Tian}, \citenamefont {Sun}, \citenamefont {Uwatoko}, \citenamefont {Wu}, \citenamefont {Chen}, \citenamefont {Wang},\ and\ \citenamefont {Cheng}}]{dongSimultaneous2024}%
  \BibitemOpen
  \bibfield  {author} {\bibinfo {author} {\bibfnamefont {Qingxin}\ \bibnamefont {Dong}}, \bibinfo {author} {\bibfnamefont {Pengtao}\ \bibnamefont {Yang}}, \bibinfo {author} {\bibfnamefont {Zhihao}\ \bibnamefont {Liu}}, \bibinfo {author} {\bibfnamefont {Yuzhi}\ \bibnamefont {Wang}}, \bibinfo {author} {\bibfnamefont {Ziyi}\ \bibnamefont {Liu}}, \bibinfo {author} {\bibfnamefont {Tong}\ \bibnamefont {Shi}}, \bibinfo {author} {\bibfnamefont {Zhaoming}\ \bibnamefont {Tian}}, \bibinfo {author} {\bibfnamefont {Jianping}\ \bibnamefont {Sun}}, \bibinfo {author} {\bibfnamefont {Yoshiya}\ \bibnamefont {Uwatoko}}, \bibinfo {author} {\bibfnamefont {Quansheng}\ \bibnamefont {Wu}}, \bibinfo {author} {\bibfnamefont {Genfu}\ \bibnamefont {Chen}}, \bibinfo {author} {\bibfnamefont {Bosen}\ \bibnamefont {Wang}}, \ and\ \bibinfo {author} {\bibfnamefont {Jinguang}\ \bibnamefont {Cheng}},\ }\href {http://arxiv.org/abs/2412.17594} {\enquote {\bibinfo {title} {Simultaneous achievement of record-breaking colossal magnetoresistance and
  angular magnetoresistance in an antiferromagnetic semiconductor {EuSe$_2$}},}\ } (\bibinfo {year} {2024}),\ \Eprint {http://arxiv.org/abs/2412.17594} {arXiv:2412.17594} \BibitemShut {NoStop}%
\bibitem [{\citenamefont {Yin}\ \emph {et~al.}(2024)\citenamefont {Yin}, \citenamefont {Shi}, \citenamefont {Liu}, \citenamefont {Kan}, \citenamefont {Qin}, \citenamefont {Feng}, \citenamefont {He}, \citenamefont {Cao}, \citenamefont {Xu}, \citenamefont {Ling}, \citenamefont {Tong}, \citenamefont {Pi},\ and\ \citenamefont {Han}}]{yinMagnetism2024}%
  \BibitemOpen
  \bibfield  {author} {\bibinfo {author} {\bibfnamefont {Huxin}\ \bibnamefont {Yin}}, \bibinfo {author} {\bibfnamefont {Xiang}\ \bibnamefont {Shi}}, \bibinfo {author} {\bibfnamefont {Xiansong}\ \bibnamefont {Liu}}, \bibinfo {author} {\bibfnamefont {Xucai}\ \bibnamefont {Kan}}, \bibinfo {author} {\bibfnamefont {Yongliang}\ \bibnamefont {Qin}}, \bibinfo {author} {\bibfnamefont {Qiyuan}\ \bibnamefont {Feng}}, \bibinfo {author} {\bibfnamefont {Miao}\ \bibnamefont {He}}, \bibinfo {author} {\bibfnamefont {Liang}\ \bibnamefont {Cao}}, \bibinfo {author} {\bibfnamefont {Hai}\ \bibnamefont {Xu}}, \bibinfo {author} {\bibfnamefont {Langsheng}\ \bibnamefont {Ling}}, \bibinfo {author} {\bibfnamefont {Wei}\ \bibnamefont {Tong}}, \bibinfo {author} {\bibfnamefont {Li}~\bibnamefont {Pi}}, \ and\ \bibinfo {author} {\bibfnamefont {Yuyan}\ \bibnamefont {Han}},\ }\bibfield  {title} {\enquote {\bibinfo {title} {Magnetism, magnetotransport properties and their correlation with magnetic field in the semiconductor-type
  {EuMnSb$_2$}},}\ }\href {https://www.sciencedirect.com/science/article/pii/S092583882401065X} {\bibfield  {journal} {\bibinfo  {journal} {Journal of Alloys and Compounds}\ }\textbf {\bibinfo {volume} {990}},\ \bibinfo {pages} {174478} (\bibinfo {year} {2024})}\BibitemShut {NoStop}%
\bibitem [{\citenamefont {Wang}\ \emph {et~al.}(2021)\citenamefont {Wang}, \citenamefont {Rogers}, \citenamefont {Yao}, \citenamefont {Nichols}, \citenamefont {Atay}, \citenamefont {Xu}, \citenamefont {Franklin}, \citenamefont {Sochnikov}, \citenamefont {Ryan}, \citenamefont {Haskel},\ and\ \citenamefont {Tafti}}]{wangColossal2021}%
  \BibitemOpen
  \bibfield  {author} {\bibinfo {author} {\bibfnamefont {Zhi-Cheng}\ \bibnamefont {Wang}}, \bibinfo {author} {\bibfnamefont {Jared~D.}\ \bibnamefont {Rogers}}, \bibinfo {author} {\bibfnamefont {Xiaohan}\ \bibnamefont {Yao}}, \bibinfo {author} {\bibfnamefont {Renee}\ \bibnamefont {Nichols}}, \bibinfo {author} {\bibfnamefont {Kemal}\ \bibnamefont {Atay}}, \bibinfo {author} {\bibfnamefont {Bochao}\ \bibnamefont {Xu}}, \bibinfo {author} {\bibfnamefont {Jacob}\ \bibnamefont {Franklin}}, \bibinfo {author} {\bibfnamefont {Ilya}\ \bibnamefont {Sochnikov}}, \bibinfo {author} {\bibfnamefont {Philip~J.}\ \bibnamefont {Ryan}}, \bibinfo {author} {\bibfnamefont {Daniel}\ \bibnamefont {Haskel}}, \ and\ \bibinfo {author} {\bibfnamefont {Fazel}\ \bibnamefont {Tafti}},\ }\bibfield  {title} {\enquote {\bibinfo {title} {Colossal magnetoresistance without mixed valence in a layered phosphide crystal},}\ }\href {https://onlinelibrary.wiley.com/doi/abs/10.1002/adma.202005755} {\bibfield  {journal} {\bibinfo  {journal} {Advanced
  Materials}\ }\textbf {\bibinfo {volume} {33}},\ \bibinfo {pages} {2005755} (\bibinfo {year} {2021})}\BibitemShut {NoStop}%
\bibitem [{\citenamefont {Krebber}\ \emph {et~al.}(2023)\citenamefont {Krebber}, \citenamefont {Kopp}, \citenamefont {Garg}, \citenamefont {Kummer}, \citenamefont {Sichelschmidt}, \citenamefont {Schulz}, \citenamefont {Poelchen}, \citenamefont {Mende}, \citenamefont {Virovets}, \citenamefont {Warawa}, \citenamefont {Thomson}, \citenamefont {Tarasov}, \citenamefont {Usachov}, \citenamefont {Vyalikh}, \citenamefont {Roskos}, \citenamefont {M{\"u}ller}, \citenamefont {Krellner},\ and\ \citenamefont {Kliemt}}]{krebberColossal2023}%
  \BibitemOpen
  \bibfield  {author} {\bibinfo {author} {\bibfnamefont {Sarah}\ \bibnamefont {Krebber}}, \bibinfo {author} {\bibfnamefont {Marvin}\ \bibnamefont {Kopp}}, \bibinfo {author} {\bibfnamefont {Charu}\ \bibnamefont {Garg}}, \bibinfo {author} {\bibfnamefont {Kurt}\ \bibnamefont {Kummer}}, \bibinfo {author} {\bibfnamefont {J{\"o}rg}\ \bibnamefont {Sichelschmidt}}, \bibinfo {author} {\bibfnamefont {Susanne}\ \bibnamefont {Schulz}}, \bibinfo {author} {\bibfnamefont {Georg}\ \bibnamefont {Poelchen}}, \bibinfo {author} {\bibfnamefont {Max}\ \bibnamefont {Mende}}, \bibinfo {author} {\bibfnamefont {Alexander~V.}\ \bibnamefont {Virovets}}, \bibinfo {author} {\bibfnamefont {Konstantin}\ \bibnamefont {Warawa}}, \bibinfo {author} {\bibfnamefont {Mark~D.}\ \bibnamefont {Thomson}}, \bibinfo {author} {\bibfnamefont {Artem~V.}\ \bibnamefont {Tarasov}}, \bibinfo {author} {\bibfnamefont {Dmitry~Yu.}\ \bibnamefont {Usachov}}, \bibinfo {author} {\bibfnamefont {Denis~V.}\ \bibnamefont {Vyalikh}}, \bibinfo {author} {\bibfnamefont
  {Hartmut~G.}\ \bibnamefont {Roskos}}, \bibinfo {author} {\bibfnamefont {Jens}\ \bibnamefont {M{\"u}ller}}, \bibinfo {author} {\bibfnamefont {Cornelius}\ \bibnamefont {Krellner}}, \ and\ \bibinfo {author} {\bibfnamefont {Kristin}\ \bibnamefont {Kliemt}},\ }\bibfield  {title} {\enquote {\bibinfo {title} {Colossal magnetoresistance in {EuZn$_2$P$_2$} and its electronic and magnetic structure},}\ }\href {https://link.aps.org/doi/10.1103/PhysRevB.108.045116} {\bibfield  {journal} {\bibinfo  {journal} {Physical Review B}\ }\textbf {\bibinfo {volume} {108}},\ \bibinfo {pages} {045116} (\bibinfo {year} {2023})}\BibitemShut {NoStop}%
\bibitem [{\citenamefont {Luo}\ \emph {et~al.}(2023)\citenamefont {Luo}, \citenamefont {Xu}, \citenamefont {Du}, \citenamefont {Yang}, \citenamefont {Chen}, \citenamefont {Cao}, \citenamefont {Song},\ and\ \citenamefont {Yuan}}]{luoColossal2023}%
  \BibitemOpen
  \bibfield  {author} {\bibinfo {author} {\bibfnamefont {Shuaishuai}\ \bibnamefont {Luo}}, \bibinfo {author} {\bibfnamefont {Yongkang}\ \bibnamefont {Xu}}, \bibinfo {author} {\bibfnamefont {Feng}\ \bibnamefont {Du}}, \bibinfo {author} {\bibfnamefont {Lin}\ \bibnamefont {Yang}}, \bibinfo {author} {\bibfnamefont {Yuxin}\ \bibnamefont {Chen}}, \bibinfo {author} {\bibfnamefont {Chao}\ \bibnamefont {Cao}}, \bibinfo {author} {\bibfnamefont {Yu}~\bibnamefont {Song}}, \ and\ \bibinfo {author} {\bibfnamefont {Huiqiu}\ \bibnamefont {Yuan}},\ }\bibfield  {title} {\enquote {\bibinfo {title} {Colossal magnetoresistance and topological phase transition in {EuZn$_2$As$_2$}},}\ }\href {https://link.aps.org/doi/10.1103/PhysRevB.108.205140} {\bibfield  {journal} {\bibinfo  {journal} {Physical Review B}\ }\textbf {\bibinfo {volume} {108}},\ \bibinfo {pages} {205140} (\bibinfo {year} {2023})}\BibitemShut {NoStop}%
\bibitem [{\citenamefont {Usachov}\ \emph {et~al.}(2024)\citenamefont {Usachov}, \citenamefont {Krebber}, \citenamefont {Bokai}, \citenamefont {Tarasov}, \citenamefont {Kopp}, \citenamefont {Garg}, \citenamefont {Virovets}, \citenamefont {M{\"u}ller}, \citenamefont {Mende}, \citenamefont {Poelchen}, \citenamefont {Vyalikh}, \citenamefont {Krellner},\ and\ \citenamefont {Kliemt}}]{usachovMagnetism2024}%
  \BibitemOpen
  \bibfield  {author} {\bibinfo {author} {\bibfnamefont {Dmitry~Yu.}\ \bibnamefont {Usachov}}, \bibinfo {author} {\bibfnamefont {Sarah}\ \bibnamefont {Krebber}}, \bibinfo {author} {\bibfnamefont {Kirill~A.}\ \bibnamefont {Bokai}}, \bibinfo {author} {\bibfnamefont {Artem~V.}\ \bibnamefont {Tarasov}}, \bibinfo {author} {\bibfnamefont {Marvin}\ \bibnamefont {Kopp}}, \bibinfo {author} {\bibfnamefont {Charu}\ \bibnamefont {Garg}}, \bibinfo {author} {\bibfnamefont {Alexander}\ \bibnamefont {Virovets}}, \bibinfo {author} {\bibfnamefont {Jens}\ \bibnamefont {M{\"u}ller}}, \bibinfo {author} {\bibfnamefont {Max}\ \bibnamefont {Mende}}, \bibinfo {author} {\bibfnamefont {Georg}\ \bibnamefont {Poelchen}}, \bibinfo {author} {\bibfnamefont {Denis~V.}\ \bibnamefont {Vyalikh}}, \bibinfo {author} {\bibfnamefont {Cornelius}\ \bibnamefont {Krellner}}, \ and\ \bibinfo {author} {\bibfnamefont {Kristin}\ \bibnamefont {Kliemt}},\ }\bibfield  {title} {\enquote {\bibinfo {title} {Magnetism, heat capacity, and electronic structure of
  {EuCd$_2$P$_2$} in view of its colossal magnetoresistance},}\ }\href {https://link.aps.org/doi/10.1103/PhysRevB.109.104421} {\bibfield  {journal} {\bibinfo  {journal} {Physical Review B}\ }\textbf {\bibinfo {volume} {109}},\ \bibinfo {pages} {104421} (\bibinfo {year} {2024})}\BibitemShut {NoStop}%
\bibitem [{\citenamefont {Zhang}\ \emph {et~al.}(2025)\citenamefont {Zhang}, \citenamefont {Fang}, \citenamefont {Weng},\ and\ \citenamefont {Wu}}]{zhanginadequacy2025}%
  \BibitemOpen
  \bibfield  {author} {\bibinfo {author} {\bibfnamefont {Shengnan}\ \bibnamefont {Zhang}}, \bibinfo {author} {\bibfnamefont {Zhong}\ \bibnamefont {Fang}}, \bibinfo {author} {\bibfnamefont {Hongming}\ \bibnamefont {Weng}}, \ and\ \bibinfo {author} {\bibfnamefont {Quansheng}\ \bibnamefont {Wu}},\ }\bibfield  {title} {\enquote {\bibinfo {title} {The inadequacy of the {{$\rho$}-T} curve for phase transitions in the presence of magnetic fields},}\ }\href {https://www.cell.com/the-innovation/abstract/S2666-6758(25)00040-2} {\bibfield  {journal} {\bibinfo  {journal} {The Innovation}\ }\textbf {\bibinfo {volume} {6}},\ \bibinfo {pages} {100837} (\bibinfo {year} {2025})}\BibitemShut {NoStop}%
\bibitem [{\citenamefont {Zhang}\ \emph {et~al.}(2022)\citenamefont {Zhang}, \citenamefont {Ni}, \citenamefont {Zhao}, \citenamefont {Hakani}, \citenamefont {Ye}, \citenamefont {DeLong}, \citenamefont {Kimchi},\ and\ \citenamefont {Cao}}]{zhangControl2022}%
  \BibitemOpen
  \bibfield  {author} {\bibinfo {author} {\bibfnamefont {Yu}~\bibnamefont {Zhang}}, \bibinfo {author} {\bibfnamefont {Yifei}\ \bibnamefont {Ni}}, \bibinfo {author} {\bibfnamefont {Hengdi}\ \bibnamefont {Zhao}}, \bibinfo {author} {\bibfnamefont {Sami}\ \bibnamefont {Hakani}}, \bibinfo {author} {\bibfnamefont {Feng}\ \bibnamefont {Ye}}, \bibinfo {author} {\bibfnamefont {Lance}\ \bibnamefont {DeLong}}, \bibinfo {author} {\bibfnamefont {Itamar}\ \bibnamefont {Kimchi}}, \ and\ \bibinfo {author} {\bibfnamefont {Gang}\ \bibnamefont {Cao}},\ }\bibfield  {title} {\enquote {\bibinfo {title} {Control of chiral orbital currents in a colossal magnetoresistance material},}\ }\href {https://www.nature.com/articles/s41586-022-05262-3} {\bibfield  {journal} {\bibinfo  {journal} {Nature}\ }\textbf {\bibinfo {volume} {611}},\ \bibinfo {pages} {467--472} (\bibinfo {year} {2022})}\BibitemShut {NoStop}%
\bibitem [{\citenamefont {Seo}\ \emph {et~al.}(2021)\citenamefont {Seo}, \citenamefont {De}, \citenamefont {Ha}, \citenamefont {Lee}, \citenamefont {Park}, \citenamefont {Park}, \citenamefont {Skourski}, \citenamefont {Choi}, \citenamefont {Kim}, \citenamefont {Cho}, \citenamefont {Yeom}, \citenamefont {Cheong}, \citenamefont {Kim}, \citenamefont {Yang}, \citenamefont {Kim},\ and\ \citenamefont {Kim}}]{seoColossal2021}%
  \BibitemOpen
  \bibfield  {author} {\bibinfo {author} {\bibfnamefont {Junho}\ \bibnamefont {Seo}}, \bibinfo {author} {\bibfnamefont {Chandan}\ \bibnamefont {De}}, \bibinfo {author} {\bibfnamefont {Hyunsoo}\ \bibnamefont {Ha}}, \bibinfo {author} {\bibfnamefont {Ji~Eun}\ \bibnamefont {Lee}}, \bibinfo {author} {\bibfnamefont {Sungyu}\ \bibnamefont {Park}}, \bibinfo {author} {\bibfnamefont {Joonbum}\ \bibnamefont {Park}}, \bibinfo {author} {\bibfnamefont {Yurii}\ \bibnamefont {Skourski}}, \bibinfo {author} {\bibfnamefont {Eun~Sang}\ \bibnamefont {Choi}}, \bibinfo {author} {\bibfnamefont {Bongjae}\ \bibnamefont {Kim}}, \bibinfo {author} {\bibfnamefont {Gil~Young}\ \bibnamefont {Cho}}, \bibinfo {author} {\bibfnamefont {Han~Woong}\ \bibnamefont {Yeom}}, \bibinfo {author} {\bibfnamefont {Sang-Wook}\ \bibnamefont {Cheong}}, \bibinfo {author} {\bibfnamefont {Jae~Hoon}\ \bibnamefont {Kim}}, \bibinfo {author} {\bibfnamefont {Bohm-Jung}\ \bibnamefont {Yang}}, \bibinfo {author} {\bibfnamefont {Kyoo}\ \bibnamefont {Kim}}, \ and\
  \bibinfo {author} {\bibfnamefont {Jun~Sung}\ \bibnamefont {Kim}},\ }\bibfield  {title} {\enquote {\bibinfo {title} {Colossal angular magnetoresistance in ferrimagnetic nodal-line semiconductors},}\ }\href {https://www.nature.com/articles/s41586-021-04028-7} {\bibfield  {journal} {\bibinfo  {journal} {Nature}\ }\textbf {\bibinfo {volume} {599}},\ \bibinfo {pages} {576--581} (\bibinfo {year} {2021})}\BibitemShut {NoStop}%
\bibitem [{\citenamefont {Liu}\ \emph {et~al.}(2024)\citenamefont {Liu}, \citenamefont {Zhang}, \citenamefont {Fang}, \citenamefont {Weng},\ and\ \citenamefont {Wu}}]{liuCombined2024}%
  \BibitemOpen
  \bibfield  {author} {\bibinfo {author} {\bibfnamefont {Zhihao}\ \bibnamefont {Liu}}, \bibinfo {author} {\bibfnamefont {Shengnan}\ \bibnamefont {Zhang}}, \bibinfo {author} {\bibfnamefont {Zhong}\ \bibnamefont {Fang}}, \bibinfo {author} {\bibfnamefont {Hongming}\ \bibnamefont {Weng}}, \ and\ \bibinfo {author} {\bibfnamefont {Quansheng}\ \bibnamefont {Wu}},\ }\bibfield  {title} {\enquote {\bibinfo {title} {Combined first-principles and boltzmann transport theory methodology for studying magnetotransport in magnetic materials},}\ }\href {https://link.aps.org/doi/10.1103/PhysRevResearch.6.043185} {\bibfield  {journal} {\bibinfo  {journal} {Physical Review Research}\ }\textbf {\bibinfo {volume} {6}},\ \bibinfo {pages} {043185} (\bibinfo {year} {2024})}\BibitemShut {NoStop}%
\bibitem [{\citenamefont {Zhang}\ \emph {et~al.}(2024{\natexlab{a}})\citenamefont {Zhang}, \citenamefont {Ni}, \citenamefont {Schlottmann}, \citenamefont {Nandkishore}, \citenamefont {DeLong},\ and\ \citenamefont {Cao}}]{zhangCurrentsensitive2024}%
  \BibitemOpen
  \bibfield  {author} {\bibinfo {author} {\bibfnamefont {Yu}~\bibnamefont {Zhang}}, \bibinfo {author} {\bibfnamefont {Yifei}\ \bibnamefont {Ni}}, \bibinfo {author} {\bibfnamefont {Pedro}\ \bibnamefont {Schlottmann}}, \bibinfo {author} {\bibfnamefont {Rahul}\ \bibnamefont {Nandkishore}}, \bibinfo {author} {\bibfnamefont {Lance~E.}\ \bibnamefont {DeLong}}, \ and\ \bibinfo {author} {\bibfnamefont {Gang}\ \bibnamefont {Cao}},\ }\bibfield  {title} {\enquote {\bibinfo {title} {Current-sensitive hall effect in a chiral-orbital-current state},}\ }\href {https://www.nature.com/articles/s41467-024-47823-2} {\bibfield  {journal} {\bibinfo  {journal} {Nature Communications}\ }\textbf {\bibinfo {volume} {15}},\ \bibinfo {pages} {3579} (\bibinfo {year} {2024}{\natexlab{a}})}\BibitemShut {NoStop}%
\bibitem [{Sup()}]{Supp}%
  \BibitemOpen
  \href@noop {} {}\bibinfo {note} {See Supplemental Material at \url{URL_will_be_inserted_by_publisher} for details of the simulation of magnetization curves, the phenomenological description of scattering enhancement near the phase transition temperature, and the simulated results for ferromagnetic system La$_{0.75}$Ca$_{0.25}$MnO$_3$}\BibitemShut {NoStop}%
\bibitem [{\citenamefont {Zhang}\ \emph {et~al.}(2024{\natexlab{b}})\citenamefont {Zhang}, \citenamefont {Li}, \citenamefont {Yang}, \citenamefont {Wei}, \citenamefont {Mi}, \citenamefont {Wang}, \citenamefont {Zhou}, \citenamefont {Yang}, \citenamefont {Chai},\ and\ \citenamefont {He}}]{zhangMagneticTransitionInduced2024}%
  \BibitemOpen
  \bibfield  {author} {\bibinfo {author} {\bibfnamefont {Yiyue}\ \bibnamefont {Zhang}}, \bibinfo {author} {\bibfnamefont {ZeYu}\ \bibnamefont {Li}}, \bibinfo {author} {\bibfnamefont {Kunya}\ \bibnamefont {Yang}}, \bibinfo {author} {\bibfnamefont {Linlin}\ \bibnamefont {Wei}}, \bibinfo {author} {\bibfnamefont {Xinrun}\ \bibnamefont {Mi}}, \bibinfo {author} {\bibfnamefont {Aifeng}\ \bibnamefont {Wang}}, \bibinfo {author} {\bibfnamefont {Xiaoyuan}\ \bibnamefont {Zhou}}, \bibinfo {author} {\bibfnamefont {Xiaolong}\ \bibnamefont {Yang}}, \bibinfo {author} {\bibfnamefont {Yisheng}\ \bibnamefont {Chai}}, \ and\ \bibinfo {author} {\bibfnamefont {Mingquan}\ \bibnamefont {He}},\ }\href {http://arxiv.org/abs/2412.01518} {\enquote {\bibinfo {title} {Magnetic-transition-induced colossal magnetoresistance in the ferrimagnetic semiconductor {Mn$_3$Si$_2$Te$_6$}},}\ } (\bibinfo {year} {2024}{\natexlab{b}}),\ \Eprint {http://arxiv.org/abs/2412.01518} {arXiv:2412.01518} \BibitemShut {NoStop}%
\bibitem [{\citenamefont {Fang}\ \emph {et~al.}(2025)\citenamefont {Fang}, \citenamefont {Hu}, \citenamefont {Chen}, \citenamefont {Liu}, \citenamefont {Yin}, \citenamefont {Ying}, \citenamefont {Wang}, \citenamefont {Wang}, \citenamefont {Li}, \citenamefont {Zhu}, \citenamefont {Xu}, \citenamefont {Pantelides},\ and\ \citenamefont {Gao}}]{fangElectrothermal2025}%
  \BibitemOpen
  \bibfield  {author} {\bibinfo {author} {\bibfnamefont {Jiaqi}\ \bibnamefont {Fang}}, \bibinfo {author} {\bibfnamefont {Jiawei}\ \bibnamefont {Hu}}, \bibinfo {author} {\bibfnamefont {Xintian}\ \bibnamefont {Chen}}, \bibinfo {author} {\bibfnamefont {Yaotian}\ \bibnamefont {Liu}}, \bibinfo {author} {\bibfnamefont {Zheng}\ \bibnamefont {Yin}}, \bibinfo {author} {\bibfnamefont {Zhe}\ \bibnamefont {Ying}}, \bibinfo {author} {\bibfnamefont {Yunhao}\ \bibnamefont {Wang}}, \bibinfo {author} {\bibfnamefont {Ziqiang}\ \bibnamefont {Wang}}, \bibinfo {author} {\bibfnamefont {Zhilin}\ \bibnamefont {Li}}, \bibinfo {author} {\bibfnamefont {Shiyu}\ \bibnamefont {Zhu}}, \bibinfo {author} {\bibfnamefont {Yang}\ \bibnamefont {Xu}}, \bibinfo {author} {\bibfnamefont {Sokrates~T.}\ \bibnamefont {Pantelides}}, \ and\ \bibinfo {author} {\bibfnamefont {Hong-Jun}\ \bibnamefont {Gao}},\ }\href {http://arxiv.org/abs/2502.11048} {\enquote {\bibinfo {title} {Electrothermal manipulation of current-induced phase transitions in
  ferrimagnetic {Mn$_3$Si$_2$Te$_6$}},}\ } (\bibinfo {year} {2025}),\ \Eprint {http://arxiv.org/abs/2502.11048} {arXiv:2502.11048} \BibitemShut {NoStop}%
\end{thebibliography}%

\end{document}